\documentclass[conference,compsoc,twoside]{IEEEtrans}
\usepackage[english]{babel}
\usepackage{caption}

\newcommand{\BTnTotalGroundTruthVisits}{{19M}}
\newcommand{\BTsimilarwebCorrelationCoefficient}{{0.83}}
\newcommand{\KFnKiwiFarmsPosts}{{10.1M}}
\newcommand{\KFnKiwiFarmsPostsRaw}{{10\,149\,700}}
\newcommand{\KFnKiwiFarmsThreads}{{48.3k}}
\newcommand{\KFnKiwiFarmsThreadsRaw}{{48\,256}}
\newcommand{\KFnKiwiFarmsActiveUsers}{{59.2k}}
\newcommand{\KFnKiwiFarmsActiveUsersRaw}{{59\,160}}
\newcommand{\KFnLolcowFarmPosts}{{4.6M}}
\newcommand{\KFnLolcowFarmPostsRaw}{{4\,593\,076}}
\newcommand{\KFnLolcowFarmThreads}{{10.0k}}
\newcommand{\KFnLolcowFarmThreadsRaw}{{10\,029}}
\newcommand{\KFnTotalPostsOfBothForum}{{14.7M}}
\newcommand{\KFnTotalPostsOfBothForumRaw}{{14\,742\,776}}

\newcommand{\KFnTotalThreadsOfBothForumRaw}{{58\,285}}
\newcommand{\KFnTelegramMessages}{{525k}}
\newcommand{\KFnTelegramReplies}{{298k}}
\newcommand{\KFnTelegramEmojis}{{356k}}

\newcommand{\KFnTelegramUsers}{{2\,502}}

\newcommand{\KFnKiwiFarmsPostsSignificanceTestStat}{{418.00}}
\newcommand{\KFnKiwiFarmsPostsSignificanceTestPValue}{{p < .0001}}

\newcommand{\KFnKiwiFarmsPostsEffectSize}{{-0.89}}

\newcommand{\KFnKiwiFarmsUsersSignificanceTestStat}{{388.00}}
\newcommand{\KFnKiwiFarmsUsersSignificanceTestPValue}{{p < .0001}}

\newcommand{\KFnKiwiFarmsUsersEffectSize}{{-0.90}}

\newcommand{\KFnKiwiFarmsThreadsSignificanceTestStat}{{416.50}}
\newcommand{\KFnKiwiFarmsThreadsSignificanceTestPValue}{{p < .0001}}

\newcommand{\KFnKiwiFarmsThreadsEffectSize}{{-0.89}}

\newcommand{\KFnTelegramMessagesSignificanceTestStat}{{7461.50}}
\newcommand{\KFnTelegramMessagesSignificanceTestPValue}{{p < .0001}}

\newcommand{\KFnTelegramMessagesEffectSize}{{0.94}}

\newcommand{\KFnTelegramRepliesSignificanceTestStat}{{6980.00}}
\newcommand{\KFnTelegramRepliesSignificanceTestPValue}{{p < .0001}}

\newcommand{\KFnTelegramRepliesEffectSize}{{0.82}}

\newcommand{\KFnTelegramEmojisSignificanceTestStat}{{7201.00}}
\newcommand{\KFnTelegramEmojisSignificanceTestPValue}{{p < .0001}}

\newcommand{\KFnTelegramEmojisEffectSize}{{0.88}}

\newcommand{\KFnTelegramUsersSignificanceTestStat}{{7361.50}}
\newcommand{\KFnTelegramUsersSignificanceTestPValue}{{p < .0001}}

\newcommand{\KFnTelegramUsersEffectSize}{{0.92}}

\newcommand{\KFnLolcowFarmPostsSignificanceTestStat}{{3349.00}}
\newcommand{\KFnLolcowFarmPostsSignificanceTestPValue}{{p = .1540}}

\newcommand{\KFnLolcowFarmPostsEffectSize}{{-0.13}}

\newcommand{\KFnLolcowFarmThreadsSignificanceTestStat}{{3830.50}}
\newcommand{\KFnLolcowFarmThreadsSignificanceTestPValue}{{p = .9791}}

\newcommand{\KFnLolcowFarmThreadsEffectSize}{{-0.00}}

\newcommand{\KFkiwifarmsAllDomainsSignificanceTestStat}{{545.00}}
\newcommand{\KFkiwifarmsAllDomainsSignificanceTestPValue}{{p < .0001}}

\newcommand{\KFkiwifarmsAllDomainsEffectSize}{{-0.86}}

\newcommand{\KFkiwifarmsnetSignificanceTestStat}{{99.00}}
\newcommand{\KFkiwifarmsnetSignificanceTestPValue}{{p < .0001}}

\newcommand{\KFkiwifarmsnetEffectSize}{{-0.97}}

\newcommand{\KFkiwifarmsruSignificanceTestStat}{{3310.50}}
\newcommand{\KFkiwifarmsruSignificanceTestPValue}{{p = .1228}}

\newcommand{\KFkiwifarmsruEffectSize}{{-0.14}}

\newcommand{\KFkiwifarmstopSignificanceTestStat}{{6125.50}}
\newcommand{\KFkiwifarmstopSignificanceTestPValue}{{p < .0001}}

\newcommand{\KFkiwifarmstopEffectSize}{{0.60}}

\newcommand{\KFkiwifarmsccSignificanceTestStat}{{3670.00}}
\newcommand{\KFkiwifarmsccSignificanceTestPValue}{{p = .6221}}

\newcommand{\KFkiwifarmsccEffectSize}{{-0.04}}

\newcommand{\KFkiwifarmsisSignificanceTestStat}{{748.00}}
\newcommand{\KFkiwifarmsisSignificanceTestPValue}{{p < .0001}}

\newcommand{\KFkiwifarmsisEffectSize}{{-0.81}}

\newcommand{\KFkiwifarmsstSignificanceTestStat}{{6560.00}}
\newcommand{\KFkiwifarmsstSignificanceTestPValue}{{p < .0001}}

\newcommand{\KFkiwifarmsstEffectSize}{{0.71}}

\newcommand{\KFnTweets}{{11\,076}}

\newcommand{\KFnPostingTweeterUsers}{{3\,886}}
\newcommand{\KFnTweeterKeyActors}{{1\,670}}
\newcommand{\KFnTweeterKeyActorsProps}{{42.97}}
\newcommand{\KFnOperatorsAnnouncements}{{107}}
\newcommand{\KFnBigFish}{{5\,159}}
\newcommand{\KFnBigFishProps}{{8.96}}
\newcommand{\KFnSmallFish}{{52\,430}}
\newcommand{\KFnSmallFishProps}{{91.04}}
\newcommand{\KFnSurvivingBigFish}{{2\,529}}
\newcommand{\KFnSurvivingBigFishProps}{{49.02}}

\newcommand{\KFnSurvivingSmallFish}{{6\,915}}
\newcommand{\KFnSurvivingSmallFishProps}{{13.19}}

\newcommand{\KFnLeavingSmallFishProps}{{86.95}}
\newcommand{\KFnReturningBigFish}{{11}}
\newcommand{\KFnReturningBigFishProps}{{0.21}}
\newcommand{\KFnReturningSmallFish}{{72}}
\newcommand{\KFnReturningSmallFishProps}{{0.14}}
\newcommand{\KFnNewFish}{{1\,488}}
\newcommand{\KFnNewUsernamesIncludingReturningFish}{{1\,571}}

\newcommand{\KFavgPostsPreDisruptionCoreSurvivors}{{1800.03}}
\newcommand{\KFavgLifetimePreDisruptionCoreSurvivors}{{1306.94}}

\newcommand{\KFavgPostsPreDisruptionCasualSurvivors}{{80.82}}
\newcommand{\KFavgLifetimePreDisruptionCasualSurvivors}{{516.84}}

\newcommand{\KFavgPostsPreDisruptionCoreSurvivorsHigherThanCasualSurvivors}{{22.3}}
\newcommand{\KFavgLifetimePreDisruptionCoreSurvivorsHigherThanCasualSurvivors}{{2.5}}
\newcommand{\KFnNodesAtTheBeginning}{{55.3k}}
\newcommand{\KFnNodesWhenTwitterCampaign}{{57.2k}}
\newcommand{\KFnNodesAtTheEnd}{{59.1k}}
\newcommand{\KFnEdgesAtTheBeginning}{{131.3M}}
\newcommand{\KFnEdgesWhenTwitterCampaign}{{137.6M}}
\newcommand{\KFnEdgesAtTheEnd}{{149.3M}}
\newcommand{\KFnKFPostsMentioningKF}{{10\,099}}
\newcommand{\KFnKFPostsMentioningKFProps}{{1.45}}
\newcommand{\KFnKFPostsMentioningCF}{{1\,515}}
\newcommand{\KFnKFPostsMentioningCFProps}{{0.22}}
\newcommand{\KFnKFPostsMentioningKFHigherThanCF}{{6.7}}
\newcommand{\KFnKFPostsMentioningBothKFCF}{{300}}
\newcommand{\KFnKFPostsMentioningBothKFCFProps}{{0.04}}
\newcommand{\KFnTelePostsMentioningKF}{{3\,794}}
\newcommand{\KFnTelePostsMentioningKFProps}{{0.72}}
\newcommand{\KFnTelePostsMentioningCF}{{286}}
\newcommand{\KFnTelePostsMentioningCFProps}{{0.05}}
\newcommand{\KFnTelePostsMentioningKFHigherThanCF}{{13.3}}
\newcommand{\KFnTelePostsMentioningBothKFCF}{{44}}
\newcommand{\KFnTelePostsMentioningBothKFCFProps}{{0.01}}
\newcommand{\KFnLFPostsMentioningKF}{{1\,494}}
\newcommand{\KFnLFPostsMentioningKFProps}{{0.31}}
\newcommand{\KFnLFPostsMentioningCF}{{197}}
\newcommand{\KFnLFPostsMentioningCFProps}{{0.04}}
\newcommand{\KFnLFPostsMentioningKFHigherThanCF}{{7.6}}
\newcommand{\KFnLFPostsMentioningBothKFCF}{{44}}
\newcommand{\KFnLFPostsMentioningBothKFCFProps}{{0.01}}

\usepackage{contour,ulem}

\contourlength{0.8pt}

\usepackage{tikz}

\usepackage{colortbl} 
\usepackage{booktabs} 
\usepackage{multirow,multicol} 
\usepackage{makecell} 
\usepackage[switch]{lineno} 
\usepackage{threeparttable} 
\usepackage{url}  
\usepackage{xspace} 
\usepackage{fancyhdr} 
\DeclareUrlCommand\myurl{\urlstyle{tt}} 
\usepackage[hidelinks,unicode,linkcolor=black]{hyperref} 

\newcommand{\snscrape}{{\small\scshape Snscrape}\xspace}
\newcommand{\snscrapemini}{{\scriptsize\scshape Snscrape}\xspace}
\newcommand{\lolcow}{{\small\scshape Lolcow}\xspace}

\newcommand{\kiwifarms}{{\small\scshape Kiwi Farms}\xspace}
\newcommand{\kiwifarmssmall}{{\footnotesize\scshape Kiwi Farms}\xspace}
\newcommand{\kiwifarmsmini}{{\scriptsize\scshape Kiwi Farms}\xspace}
\newcommand{\lolcowfarms}{{\small\scshape Lolcow Farm}\xspace}
\newcommand{\lolcowfarmsmall}{{\footnotesize\scshape Lolcow Farm}\xspace}
\newcommand{\lolcowfarmsmini}{{\scriptsize\scshape Lolcow Farm}\xspace}
\newcommand{\silkroad}{{\small\scshape Silk Road}\xspace}
\newcommand{\hansamarket}{{\small\scshape Hansa Market}\xspace}
\newcommand{\raidforums}{{\small\scshape Raid Forums}\xspace}
\newcommand{\breachforums}{{\small\scshape Breach Forums}\xspace}
\newcommand{\exposedforums}{{\small\scshape Exposed Forums}\xspace}
\newcommand{\onniforums}{{\small\scshape Onni Forums}\xspace}
\newcommand{\dailystormer}{{\small\scshape Daily Stormer}\xspace}
\newcommand{\eightchan}{{\small\scshape 8Chan}\xspace}
\newcommand{\extremebb}{{\small\scshape ExtremeBB}\xspace}
\newcommand{\covid}{{\small\scshape Covid-19}\xspace}

\newcommand{\ccc}{Cambridge Cybercrime Centre}

\newcommand{\hcaptcha}{hCaptcha\xspace}
\newcommand{\kiwiflare}{KiwiFlare\xspace}
\newcommand{\cloudflare}{Cloudflare\xspace}
\newcommand{\cloudflaresmall}{Cloudflare\xspace}
\newcommand{\ddosguard}{DDoS-Guard\xspace}
\newcommand{\diamwall}{DiamWall\xspace}
\newcommand{\harica}{Harica\xspace}

\IEEEoverridecommandlockouts 

\begin{document}
\bstctlcite{IEEEtran:BSTcontrol}

\pagestyle{fancy}
\fancyhead[RO,LE]{\small{In Proceedings of the IEEE Symposium on Security and Privacy 2024\vspace{0.05mm}}}
\fancyfoot[LO,RE]{\small{Anh V. Vu, Alice Hutchings, and Ross Anderson}}
\fancyfoot[RO,LE]{\small{\thepage}}
\fancyfoot[CO,CE]{}

\title{No Easy Way Out: the Effectiveness of Deplatforming an Extremist\\Forum to Suppress Hate and Harassment}
\author{\IEEEauthorblockN{Anh V. Vu}
\IEEEauthorblockA{University of Cambridge\\Cambridge Cybercrime Centre\\anh.vu@cl.cam.ac.uk}
\and
\IEEEauthorblockN{Alice Hutchings}
\IEEEauthorblockA{University of Cambridge\\Cambridge Cybercrime Centre\\alice.hutchings@cl.cam.ac.uk}
\and
\IEEEauthorblockN{Ross Anderson}
\IEEEauthorblockA{University of Cambridge\\and University of Edinburgh\\ross.anderson@cl.cam.ac.uk}}
\maketitle

\begin{abstract}
Legislators and policymakers worldwide are debating options for suppressing illegal, harmful and undesirable material online. Drawing on several quantitative data sources, we show that deplatforming an active community to suppress online hate and harassment, even with a substantial concerted effort involving several tech firms, can be hard. Our case study is the disruption of the largest and longest-running harassment forum \kiwifarmssmall in late 2022, which is probably the most extensive industry effort to date. Despite the active participation of a number of tech companies over several consecutive months, this campaign failed to shut down the forum and remove its objectionable content. While briefly raising public awareness, it led to rapid platform displacement and traffic fragmentation. Part of the activity decamped to Telegram, while traffic shifted from the primary domain to previously abandoned alternatives. The forum experienced intermittent outages for several weeks, after which the community leading the campaign lost interest, traffic was directed back to the main domain, users quickly returned, and the forum was back online and became even more connected. The forum members themselves stopped discussing the incident shortly thereafter, and the net effect was that forum activity, active users, threads, posts and traffic were all cut by about half. The disruption largely affected casual users (of whom roughly 87\% left), while half the core members remained engaged. It also drew many newcomers, who exhibited increasing levels of toxicity during the first few weeks of participation. Deplatforming a community without a court order raises philosophical issues about censorship versus free speech; ethical and legal issues about the role of industry in online content moderation; and practical issues on the efficacy of private-sector versus government action. Deplatforming a dispersed community using a series of court orders against individual service providers appears unlikely to be very effective if the censor cannot incapacitate the key maintainers, whether by arresting them, enjoining them or otherwise deterring them.
\end{abstract}

\section{Introduction}
\noindent Online content is now prevalent, widely accessible, and influential in shaping public discourse. Yet while online places facilitate free speech, they do the same for hate speech~\cite{mondal2017measurement}, and the line between the two is often contested. Some cases of stalking, bullying, and doxxing such as Gamergate have had real-world consequences, including violent crime and political mobilisation~\cite{aghazadeh2018gamergate}. Content moderation has become a critical function of tech companies, but also a political tussle space, since abusive accounts may affect online communities in significantly different ways~\cite{kumar2023understanding}. Online social platforms employ various mechanisms, for example, artificial intelligence~\cite{gunton2022use}, to detect, moderate, and suppress objectionable content~\cite{singhal2023sok}, including ``hard'' and ``soft'' techniques~\cite{de2021deplatforming}. These range from reporting users of illegal content to the police, through deplatforming users breaking terms of service~\cite{rogers2020deplatforming}, to moderating legal but obnoxious content~\cite{habib2019act}, which may involve actions such as flagging it with warnings, downranking it in recommendation algorithms~\cite{gillespie2022not}, or preventing its being monetised through ads~\cite{kayes2015social}.

Deplatforming may mean blocking individual users, but sometimes the target is not a single bad actor, but a whole community, such as one involved in crime~\cite{hutchings2016taking}. It can be undertaken by industry, as when \cloudflare, GoDaddy, Google and some other firms terminated service for the \dailystormer after the Unite the Right rally in Virginia in 2017~\cite{cloudflaretakedowndailystormer} and for \eightchan in August 2019~\cite{cloudflaretakedown8chan}; or by law enforcement, as with the FBI taking down DDoS-for-hire services in 2018~\cite{collier2019booting,kopp2019ddos} and 2022~\cite{fbibooterseizure2022first,fbibooterseizure2022second}, and seizing \raidforums in 2022~\cite{raidforumtakedown}. Industry disruption has often been short-lived; both \eightchan and \dailystormer re-emerged or relocated shortly after being disrupted. Police intervention is often slow and less effective, and its impact may also be temporary~\cite{hutchings2016taking}. After the FBI terminated \silkroad~\cite{silkroadtakedown}, the online drug market fragmented among multiple smaller ones~\cite{soska2015measuring}. The seizure of \raidforums~\cite{raidforumtakedown} led to the emergence of its successors \breachforums, \exposedforums, and \onniforums. Furthermore, the FBI takedowns of DDoS-for-hire services cut the attack volume significantly, yet the market recovered rapidly~\cite{collier2019booting,kopp2019ddos}. 

\kiwifarms is the largest and longest-running online harassment forum~\cite{pless2016kiwi}. It is often associated with real-life trolling and doxxing campaigns against feminists, gay rights campaigners and minorities such as disabled, transgender, and autistic individuals; some have killed themselves after being harassed~\cite{ambreen2019fyi}. Despite being unpleasant and widely controversial, the forum has been online for a decade and had been shielded by \cloudflare's DDoS protection for years. This came to an end following serious harassment by forum members of a Canadian trans activist, culminating in a swatting incident in August 2022.\footnote{This is when a harasser falsely reports a violent crime in progress at the victim's home, resulting in the arrival of a special-weapons-and-tactics (SWAT) team to storm the premises, placing the victim and family at risk.} This resulted in a community-led campaign on Twitter to pressure \cloudflare and other tech firms to drop the forum~\cite{wiredkiwifarms}. This escalated quickly, generating significant social media attention and mainstream headlines. A series of tech firms then attempted to take the forum down; they included DDoS protection services, infrastructure providers, and even some Tier-1 networks~\cite{cloudflarebloackskiwifarms,kiwifarmsblockedbyddosguard,kiwifarmsblockedbydiamwall,kiwifarmsblockedtopdomain}. This extraordinary series of events lasted for a few months and was the most sustained effort to date to suppress an active online hate community. It is notable that tech firms gave in to public pressure in this case, while they have in the past resisted substantial pressure from governments. 

Existing studies have investigated the efficacy of deplatforming social-media users~\cite{jhaver2021evaluating,chandrasekharan2017you,saleem2018aftermath,innes2023platforming,rauchfleisch2021deplatforming,ali2021understanding,bryanov2022other}, yet there has been limited research -- both quantitative and qualitative -- into the effectiveness of industry disruptions against standalone hate communities such as bulletin-board forums, which tend to be more resilient as the content can be fully backed up and restored by the admins. This paper investigates how well the industry -- the entities offering digital infrastructure for online services such as hosting and domain providers, security and protection services, certificate authorities, and ISP networks -- dealt with a hate and harassment site.

We outline the disruption landscape in~\S\ref{sec:background}, then describe our methods, datasets, and ethics in~\S\ref{sec:methods-and-datasets}. Our ultimate goal is to evaluate the efficacy of the effort, and to understand the impacts and challenges of deplatforming as a means to suppress online hate and harassment. Our primary research questions are tackled in subsequent sections: the impact of deplatforming on the forum activity and traffic is assessed in~\S\ref{sec:forum-activity-and-displacement}; the changes in the behaviour of forum members when their gathering place is disrupted, as well as the effects on the forum operators and the community who started the campaign are examined in~\S\ref{sec:relevant-stakeholders}. We discuss the role of industry in tackling online harassment, censorship and content regulation, as well as legal, ethical, and policy implications of the incident in~\S\ref{sec:discussion}. Our data collection and analyses were approved by our institutional Ethics Review Board (ERB). Our data and scripts are available to academics on request. 

\begin{figure}[t]
    \centering
    \includegraphics[width=0.48\textwidth]{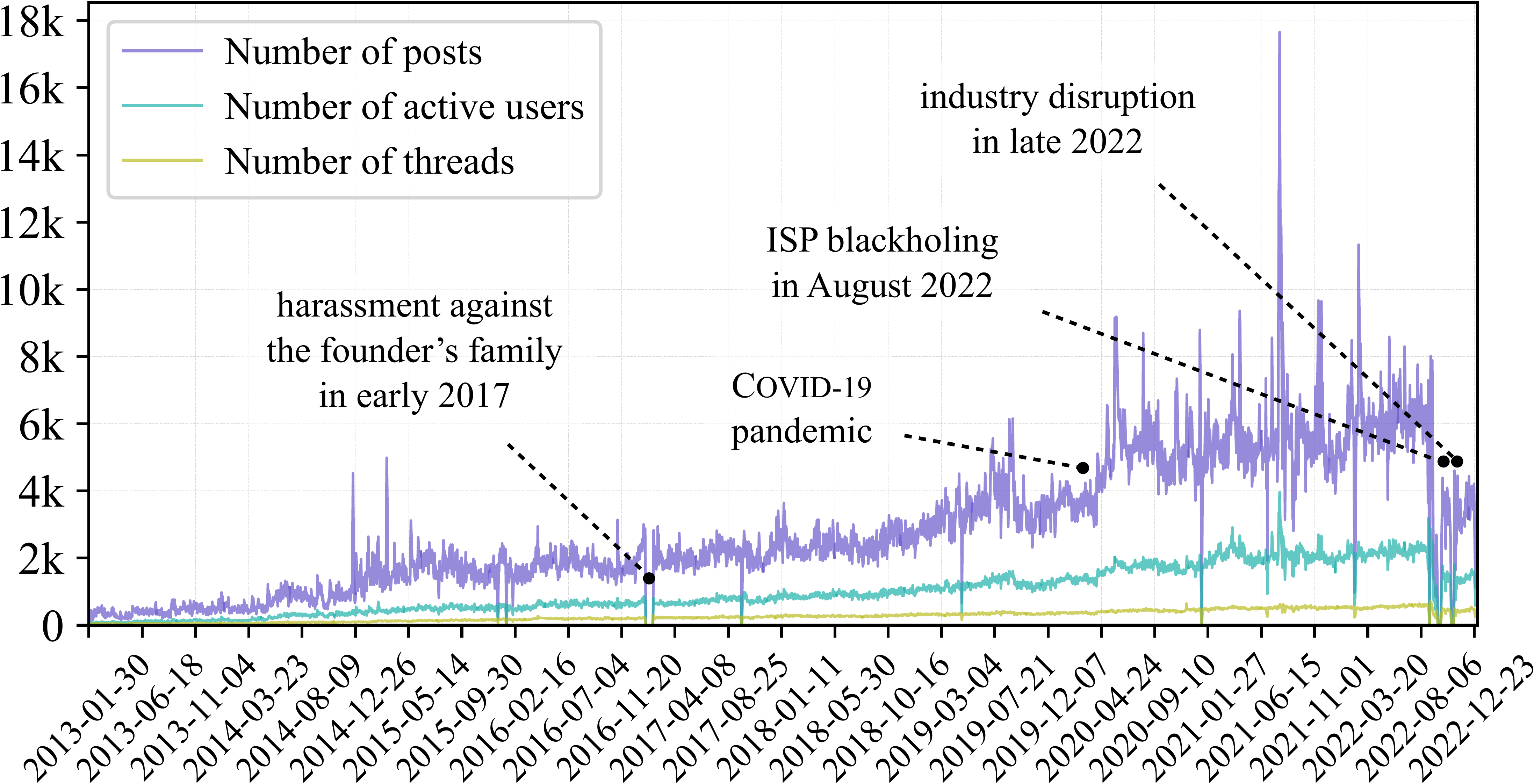}
    \caption{Number of daily posts, threads, users; and the incidents affecting \kiwifarms during its one-decade lifetime.}
    \label{fig:kiwifarms-activity-lifetime}
\end{figure}
\begin{figure*}[t]
    \centering
    \includegraphics[width=\textwidth]{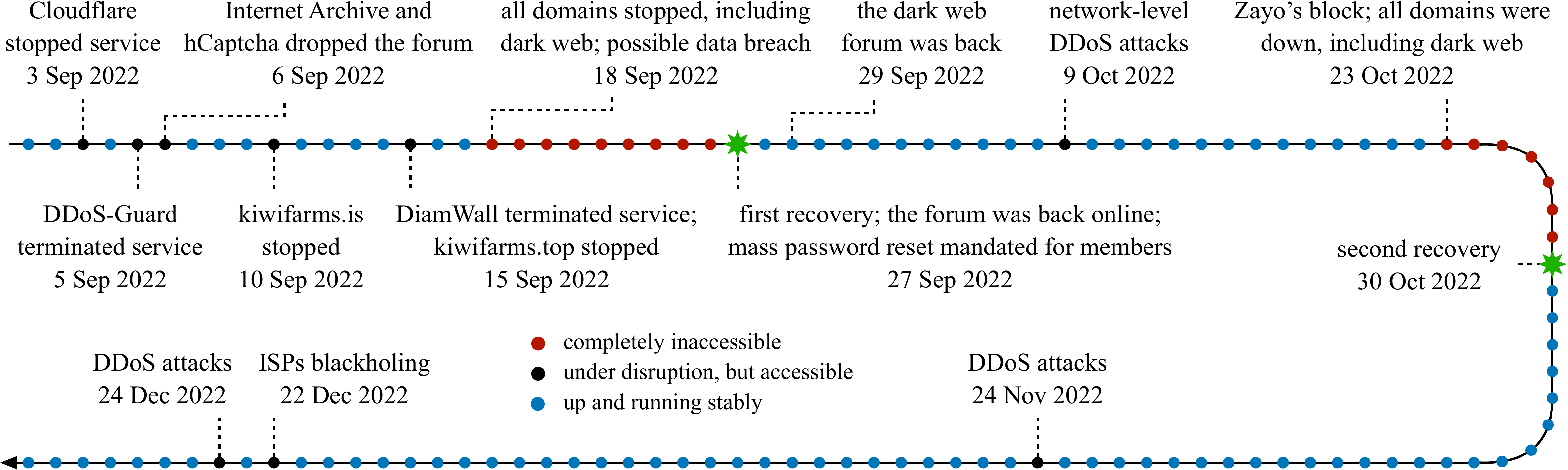}\vspace{2mm}
    \caption{Major incidents disrupting \kiwifarms from September to December 2022. Green stars indicate the forum recovery.}
    \label{fig:disruption-timeline}
\end{figure*}
 
\section{Deplatforming and the Impacts} \label{sec:background}
\noindent There is a complex ecosystem of online abuse that has been evolving for decades~\cite{thomas2021sok}, where toxic content, surveillance, and content leakage are growing threats~\cite{wei2023there,aliapoulios2021large}. While the number of personally targeted victims is relatively low, an increasing number of individuals, including children, are being exposed to online hate speech~\cite{williams2019hatred}. There can be a large grey area between criminal behaviour and socially acceptable behaviour online, just as in real life. And just as a pub landlord will throw out rowdy customers, so platforms have acceptable-use policies backed by content moderation~\cite{pater2016characterizations}, to enhance the user experience and protect advertising revenue~\cite{jimenez2021economics}. 

Deplatforming refers to blocking, excluding or restricting individuals or groups from using online services, on the grounds that their activities are unlawful, or that they do not comply with the platform's acceptable-use policy~\cite{rogers2020deplatforming}. Various extremists and criminals have been exploiting online platforms for over thirty years, resulting in a complex ecosystem in which some harms are prohibited by the criminal law (such as terrorist radicalisation and child sex abuse material) while many others are blocked by platforms seeking to provide welcoming spaces for their users and advertisers. For a history and summary of current US legislative tussles and their possible side-effects, see Fishman~\cite{fishman2023dual}. The idea is that if a platform is used to disseminate abusive speech, removing the speech or indeed the speakers could restrict its spread, make it harder for hate groups to recruit, organise and coordinate, and ultimately protect individuals from mental and physical harm. Deplatforming can be done in various ways, ranging from limiting users' access and restricting their activity for a time period, to suspending an account, or even stopping an entire group of users from using one or more services. For example, groups banned from major platforms can displace to other channels, whether smaller websites or messenger services~\cite{rogers2020deplatforming}. 

Different countries draw the line between free speech and hate speech differently. For example, the USA allows the display of Nazi symbols while France and Germany do not~\cite{schauer2005exceptional}. Private firms offering ad-supported social networks generally operate much more restrictive rules, as their advertisers do not want their ads appearing alongside content that prospective customers are likely to find offensive. People wishing to generate and share such material therefore tend to congregate on smaller forums. Some argue that taking down such forums infringes on free speech and may lead to censorship of legitimate voices and dissenting opinions, especially if it is perceived as politically motivated. Others maintain that deplatforming is necessary to protect vulnerable communities from harm. Debates rage in multiple legislatures; as one example, the UK Online Safety Bill will enable the (politically-appointed) head of Ofcom, the UK broadcast regulator, to obtain court orders to shut down online places that are considered harmful~\cite{anderson2023onlinesafetybill}. This lead us to ask: how effective might such an order be?

\subsection{Related Work} 
\noindent Most studies assessing the impact of deplatforming have worked with data on social networks. Deplatforming users may reduce activity and toxicity levels of relevant actors on Twitter~\cite{jhaver2021evaluating} and Reddit~\cite{chandrasekharan2017you,saleem2018aftermath}, limit the spread of conspiratorial disinformation on Facebook~\cite{innes2023platforming}, reduce the engagement of peripheral members with hateful content~\cite{thomas2023disrupting}, and minimise disinformation and extreme speech on YouTube~\cite{rauchfleisch2021deplatforming}. But deplatforming has often made hate groups and individuals even more extreme, toxic and radicalised. They may view the disruption of their platform as an attack on their shared beliefs and values, and move to even more toxic places to continue spreading their message. There are many examples: the Reddit ban of r/incels in November 2017 led to the emergence of two standalone forums, incels.is and incels.net, which then grew rapidly; users banned from Twitter and Reddit exhibit higher levels of toxicity when migrating to Gab~\cite{ali2021understanding}; users migrated to their own standalone websites after getting banned from r/The\_Donald expressed higher levels of toxicity and radicalisation, even though their posting activity on the new platform decreased~\cite{horta2021platform,russo2023spillover}; the `Great Deplatforming' directed users to other less regulated, more extreme platforms~\cite{buntain2023cross}; the activity of many right-wing users moved to Telegram increased multi-fold after being banned on major social media~\cite{bryanov2022other}; users banned from Twitter are more active on Gettr~\cite{mekacher2023systemic}; communities migrated to Voat from Reddit can be more resilient~\cite{monti2023online}; and roughly half of QAnon users moved to Poal after the Voat shutdown~\cite{papasavva2023waiting}. Blocking can also be ineffective for technical and implementation reasons: removing Facebook content after a delay appears to have been ineffective and had limited impact due to the short cycle of users' engagement~\cite{goldstein2023understanding}. 

The major limitation of focusing on social networks is that these platforms are often under the control of a single tech company and thus content can be permanently removed without effective backup and recovery. We instead examine deplatforming a standalone website involving a concerted effort on a much wider scale by a series of tech companies, including some big entities that handle a large amount of Internet traffic. Such standalone communities, for instance, websites and forums, may be more resilient as the admin has control of all the content, facilitating easy backups and restores. While existing studies measure changes in posting activity and the behaviours of actors when their place is disrupted, we also provide insights about other stakeholders such as the forum operators, the community leading the campaign, and the tech firms that attempted the takedown.

Previous work has documented the impacts of law enforcement and industry interventions on online cybercrime marketplaces~\cite{soska2015measuring}, cryptocurrency market price~\cite{abramova2021out}, DDoS-for-hire services~\cite{collier2019booting,kopp2019ddos}, the Kelihos, Zeus, and Nitol botnets~\cite{nadji2013beheading}, and the well-known click fraud network ZeroAccess~\cite{pearce2014characterizing}; yet how effective a concerted effort of several tech firms can be in deplatforming an extreme and radicalised community remains unstudied.

\subsection{The Kiwi Farms Disruption}
\noindent \kiwifarms had been growing steadily over a decade (see Figure~\ref{fig:kiwifarms-activity-lifetime}) and had been under \cloudflare's DDoS protection for some years.\footnote{\cloudflaresmall's service tries to detect suspicious patterns and drop malicious ones, only letting legitimate requests through.} An increase of roughly 50\% in forum activity happened during the \covid lockdown starting in March 2020, presumably as people were spending more time online. Prior interventions have resulted in the forum getting banned from Google Adsense, and from Mastercard, Visa and PayPal in 2016; from hundreds of VPS providers between 2014--2019~\cite{kiwifarmsprinciple}; and from selling merchandise on the print-on-demand marketplace Redbubble in 2016. XenForo, a close-source forum platform, revoked its license in late 2021~\cite{kiwifarmsxenforo}. DreamHost stopped its domain registration in July 2021 after a software developer killed himself after being harassed by the site's users. This did not disrupt the forum as it was given 14 days to seek another registrar~\cite{dreamhoststop}. While these interventions may have had negative effects on its profit and loss account, they did not impact its activity overall. The only significant disruption in the forum's history was between 22 January and 9 February 2017 (19 days), when the forum's owner suspended it himself due to his family being harassed~\cite{kfdownharrasmentfamily}.\footnote{Minor suspensions observed in our forum dataset are on 2 Feb 2013, 24 Jan 2016, 29 Sep 2017, and 11 Jan 2021, yet without any clear reasons.}

The disruption studied in this work was started by the online community in 2022. A malicious alarm was sent to the police in London, Ontario by a forum member on 5 August 2022, claiming that a Canadian trans activist had committed murders and was planning more, leading to her being swatted~\cite{wiredkiwifarms}. She and her family were then repeatedly tracked, doxxed, threatened, and generally harassed. In return, she launched a campaign on Twitter on 22 August 2022 under the hashtag \#dropkiwifarms and planned a protest outside \cloudflare's headquarters to pressure the company to deplatform the site~\cite{keffalsledprotestcloudflare}. This campaign generated lots of attention and mainstream headlines, which ultimately resulted in several tech firms trying to shut down the forum. This is the first time that the forum was completely inaccessible for an extended period due to an external action, with no activity on any online places including the dark web. It attempted to recover twice, but even when it eventually returned online, the overall activity was roughly halved.

The majority of actions taken to disrupt the forum occurred within the first two months of the campaign. Most of them were widely covered in the media and can be checked against public statements made by the industry and the forum admins' announcements (see Figure~\ref{fig:disruption-timeline}). The forum came under a large DDoS attack on 23 August 2022, one day after the campaign started. It was then unavailable from 27 to 28 August 2022 due to ISP blackholing. \cloudflare terminated their DDoS prevention service on 3 September 2022 -- just 12 days after the Twitter campaign started -- due to an ``unprecedented emergency and immediate threat to human life''~\cite{cloudflarebloackskiwifarms}. The forum was still supported by \ddosguard (a Russian competitor to \cloudflare), but that firm also suspended service on 5 September 2022~\cite{kiwifarmsblockedbyddosguard}. The forum was still active on the dark web but this .onion site soon became inaccessible too. On 6 September 2022, \hcaptcha dropped support; the forum was removed from the Internet Archive on the same day~\cite{kiwifarmsdroppedbyinternetarchive}. This left it under \diamwall's DDoS protection and hosted on VanwaTech -- a hosting provider describing themselves as neutral and non-censored~\cite{kiwifarmsvanwatech}. On 15 September 2022, \diamwall terminated their protection~\cite{kiwifarmsblockedbydiamwall} and the `.top' domain provider also stopped support~\cite{kiwifarmsblockedtopdomain}. The forum was completely down from 19 to 26 September 2022 and from 23 to 29 October 2022. From 23 October 2022 onwards, several ISPs intermittently rejected announcements or blackholed routes to the forum due to violations of their acceptable use policy, including Voxility and Tier-1 providers such as Lumen, Arelion, GTT and Zayo. This is remarkable as there are only about 15 Tier-1 ISPs in the world. The forum admin devoted extensive effort to maintaining the infrastructure, fixing bugs, and providing guidance to users in response to password breaches. Eventually, by routing through other ISPs, \kiwifarms was able to get back online on the clearnet and remain stable, particularly following its second recovery in October 2022.

\begin{table}[t]
\centering
\small
\caption{Complete snapshots of public posts on \kiwifarms and its primary competitor \lolcowfarms until 31 Dec 2022.}
\setlength{\tabcolsep}{0.675em}
\begin{tabular}{lrrr}
\toprule
\small{Forums} & No. posts & No. threads & No. active users \\
\midrule
\kiwifarmssmall & \KFnKiwiFarmsPostsRaw & \KFnKiwiFarmsThreadsRaw & \KFnKiwiFarmsActiveUsersRaw\\
\lolcowfarmsmall & \KFnLolcowFarmPostsRaw & \KFnLolcowFarmThreadsRaw & Unavailable\\
\midrule
Total & \KFnTotalPostsOfBothForumRaw & \KFnTotalThreadsOfBothForumRaw & Unavailable\\
\bottomrule
\end{tabular}
\label{tab:dataset-taxonomy}
\end{table}
\section{Methods, Datasets, and Ethics} \label{sec:methods-and-datasets}
\noindent Our primary method is data-driven, with findings supported by quantitative evidence derived from multiple longitudinal data sources, which we collect on a regular basis. Where quantitative measurements require enrichment -- as when analysing relevant public statements of tech firms directly involved in the disruption, and announcements made by the forum operators -- we use qualitative content analysis.

\subsection{Forum and Imageboard Discussions} 
\noindent Besides common mainstream social media channels like Facebook and Twitter, independent platforms such as xenForo\footnote{The xenForo Platform: https://xenforo.com/} and Infinity\footnote{The Infinity Imageboard: https://github.com/ctrlcctrlv/infinity/} have gained popularity as tools for building online communities. Despite being less visible and requiring more upkeep, these can offer greater resistance against external intervention as the operators have full control over the content and databases, thereby allowing easy backup and redeployment in case of disruption. These platforms typically share a hierarchical data structure ranging from bulletin boards down to threads linked to specific topics, each containing several posts. While facilitating free speech, these also increasingly nurture and disseminate hate and abusive speech. We have been scraping the two most active forums associated with online harassment for years due to their increasingly toxic content, as part of the \extremebb dataset~\cite{vu2023extremebb}: \kiwifarms and \lolcowfarms.

Our collection includes not only posts but also associated metadata such as posting time, user profiles, reactions, and levels of \textit{toxicity}, \textit{identity attack} and \textit{threat} measured by the Google Perspective API as of January 2023.\footnote{Google Perspective API: https://perspectiveapi.com/} Perspective API also offers other measures such as \textit{insult} and \textit{profanity}~\cite{perspectiveapiconcepts}, but we exclude these due to lack of relevance to the aim of this paper. This API uses crowdsourced annotations for model training and substantially outperforms the alternatives~\cite{zannettou2020measuring}. We strive to ensure data completeness by designing our scrapers to visit all sub-forums, threads, and posts while keeping track of every single crawl's progress to resume incrementally in case of any interruption. A summary of the forum discussion data is shown in Table~\ref{tab:dataset-taxonomy}.

\kiwifarms is built on xenForo, but the operators have been maintaining the forum by their own efforts since late 2021 when xenForo officially revoked their license. Our data covers the entire history of the forum from early January 2013 to the end of 2022 with \KFnKiwiFarmsPosts~posts in \KFnKiwiFarmsThreads~threads made by \KFnKiwiFarmsActiveUsers~active users, providing a full landscape through its evolution over time. While some extremist forums experienced fluctuating activity and rapid declines in recent years~\cite{vu2023extremebb}, \kiwifarms has shown stable growth until being significantly disrupted in 2022 (see Figure~\ref{fig:kiwifarms-activity-lifetime}). Our data precisely capture major reported suspensions, including those in 2017 and 2022.

The primary rival of \kiwifarms is \lolcowfarms, an imageboard built on Infinity~\cite{kfcompetitorsimilarweb,kfcompetitorsemrush}. While \kiwifarms discussions are largely text-based, \lolcowfarms is centred on descriptive images. While \kiwifarms users adopt pseudonyms, \lolcowfarms users mostly remain hidden under the unified `Anonymous' handle. We gathered a complete snapshot of \lolcowfarms from its inception in June 2014 to the end of 2022, encompassing \KFnLolcowFarmPosts~posts made in \KFnLolcowFarmThreads~threads. \lolcowfarms has much fewer threads, but each typically contains lots of posts. This collection brings the total number of posts for both forums to \KFnTotalPostsOfBothForum~(and still growing). We exclude \lolcow, a smaller competitor to \kiwifarms (also based on xenForo), as it vanished in mid-2022 and had less than 30k posts in total. As \lolcowfarms is now the largest competitor, analysing it lets us estimate platform displacement when \kiwifarms was down.

\subsection{Telegram Chats}
\noindent During periods of inaccessibility, the activity level increased in the Telegram groups associated with \kiwifarms. There are two channels: one is primarily used by the forum operators to disseminate announcements and updates, particularly about where and when the forum could be accessed; and one is adopted by the forum users mainly for normal discussions. Both channels permit public access, allowing people to join and view historical messages. We used Telethon\footnote{Telethon: https://telethon.dev/} to collect a snapshot of these channels during their entire lifespan until the end of 2022, encompassing \KFnTelegramMessages~messages, \KFnTelegramReplies~replies, and associated metadata such as view counts and \KFnTelegramEmojis~emoji reactions made by \KFnTelegramUsers~active users. The data is likely complete as our scraper is running in near real time, and messages with metadata are fully captured through the use of official Telegram APIs. As the forum operators are highly incentivised to keep users quickly informed, their announcements provide a reliable incident and response timeline.

\subsection{Web Traffic and Search Trends Analytics} 
\noindent We found from announcements in the Telegram group that \kiwifarms could be accessed through six major domains: the primary one is \url{kiwifarms.net} and four alternatives are \url{kiwifarms.ru}, \url{kiwifarms.top}, \url{kiwifarms.is}, and \url{kiwifarms.st}, while a Pleroma decentralised web version is at \url{kiwifarms.cc}.\footnote{Other domains include \url{kiwifarms.tw} and \url{kiwifarms.hk}, however they are either new or insignificant so their traffic data is trivial.} To investigate how users navigated across these domains when the forum experienced disruption, we analysed traffic analytics towards all six domains provided by Similarweb -- the leading platform in the market providing insights and intelligence into web traffic and performance.\footnote{Similarweb: https://similarweb.com/. Another popular web analytics is Semrush at https://semrush.com/, but it does not offer daily statistics.} Their reports aggregate anonymous statistics from multiple inputs, including their own analytic services, data sharing from ISPs and other measurement companies, data crawled from billions of websites, and device traffic data (both website and app) such as plugins, add-ons and pixel tracking. Their algorithm then extrapolates the substantial aggregated data to the entire Internet space. Their estimation therefore may not be completely precise, but reliably reflects trends at both global and country levels. To test that reliability, we deployed our own infrastructure to collect over \BTnTotalGroundTruthVisits~ground-truth traffic records over six months, grouped them into 30-minute sessions then compared with Similarweb visits. We find that while underestimating the amount of traffic due to how repeat pageviews are counted, Similarweb is able to capture trends with a strong positive linear relationship (Pearson correlation coefficient $r=\BTsimilarwebCorrelationCoefficient$). Our analysis in the next section also suggests a high correlation between the traffic data and the forum activity.

As Similarweb does not offer an academic license, we use a free trial account\footnote{A business subscription offers 6 months of historical data, but neither it nor the free trial provides access to longitudinal country-based records.} to access longitudinal web traffic and engagement data going back the past three months. This includes information about total visits, unique visitors, visit duration, pages per visit, bounce rate, and page views. It also provides figures on search activity, data for marketing such as visit sources (e.g., direct, search, email, social, referral, ads), and non-temporal insight into audience geography and demographics. These data, covering both desktop and mobile traffic, provide valuable perspectives. They span from July to December 2022, two months before and four months after the disruption; this time frame is sufficient as there was no significant industry intervention against the forum in the past (as shown in Figure~\ref{fig:kiwifarms-activity-lifetime}), and the disruption campaign mostly ended after a few months (see~\S\ref{sec:forum-activity-and-displacement}). In addition, we also collected search trends by countries and territories over time from Google Trends, covering the entire lifetime of the forum. Both of these datasets are likely to be complete as they were gathered directly from Similarweb and Google.

\subsection{Tweets Made by the Online Community}
\noindent The disruption campaign started on Twitter on 22 August 2022 with tweets posted under the hashtag \#dropkiwifarms. We gathered the main tweets plus associated metadata, such as posting time and reactions (e.g., replies, retweets, likes, and quotes) using \snscrape, an open-source Python framework for social network scrapers.\footnote{\snscrapemini: https://github.com/JustAnotherArchivist/snscrape/} As they use Twitter APIs as the underlying method, the data are likely to be complete. We collected \KFnTweets~tweets made by \KFnPostingTweeterUsers~users, spanning the entire campaign period. This data helps us understand the community reaction throughout the campaign, when the industry took action, and when the forum recovered. There might be more related tweets without the hashtag \#dropkiwifarms of which we are unaware, but scanning the whole Twitter space is infeasible. It is likely that the trend measured by our collection is representative as the campaign was congregated around this hashtag.

\subsection{Data Licensing} \label{subsec:data-licensing}
\noindent Our datasets and scripts for data collection and analysis are available to academics, as well as an interactive web portal to assist those who lack technical skills to access our data~\cite{pete2022postcog}. However, as both researchers and actors such as forum members might be exposed to risk and harm~\cite{doerfler2021m}, we decline to make our data publicly accessible. It is our standard practice at the \ccc~to require our licensees to sign an agreement to prevent misuse, to ensure the data will be handled appropriately, and to keep us informed about research outcomes~\cite{wilson2023identifying}. We have a long history of sharing such sensitive data, and robust procedures carefully crafted in conjunction with legal academics, university lawyers and specialist external counsel to enable data sharing across multiple jurisdictions.

\subsection{Ethical Considerations} \label{subsec::ethics}
\noindent Our work was formally approved by our institutional Ethics Review Board (ERB) for data collection and analysis. Our datasets are collected on publicly available forums and channels, which are accessible to all. We collected the forum when it was hosted in the US; according to a 2022 US court case, scraping public data is legal~\cite{webscrapinglegal}. Our scraping method does not violate any regulations and does not cause negative consequences to the targeted websites e.g., bandwidth congestion or denial of service. It would be impractical to send thousands of messages to gain consent from all forum and Telegram members; we assume they are aware that their activity on public online places will be widely accessible.

In contrast to some previous work on online forums, we name the investigated forums in this paper. Pseudonymising the forum name is pointless because of the high-profile campaign being studied. Thus, we avoid the pretence that the forum is not identifiable and shift the focus to accounting for the potential harms to both researchers and involved actors associated with our research. We designed our analysis to operate ethically and collectively by only presenting aggregated behaviours to avoid private and sensitive information of individuals being inferred. This is in accordance with the British Society of Criminology Statement on Ethics~\cite{britishethics}. 

Researchers may be at risk and may experience various elevated digital threats when doing work on sensitive resources~\cite{doerfler2021m,warford2022sok}. Studying extremist forums may introduce a higher risk of retaliation than other forums, resulting in mental or physical harm. We have taken measures to minimise potential harm to researchers and involved actors when doing studies with human subjects and at-risk populations~\cite{marwick2016best,bhalerao2022ethical}. For example, we consider options to anonymise authors' names or use pseudonyms for any publication related to the project, including this paper, if necessary. We also refrain from directly looking at media, which may cause emotional harm; our scrapers thus only collect text while discarding images and videos. Although all datasets are widely accessible and can be gathered by the public, we refrain from scraping private and protected posts behind the login wall due to safety and legality concerns.

\begin{figure}[t]
    \centering
    \includegraphics[width=0.48\textwidth]{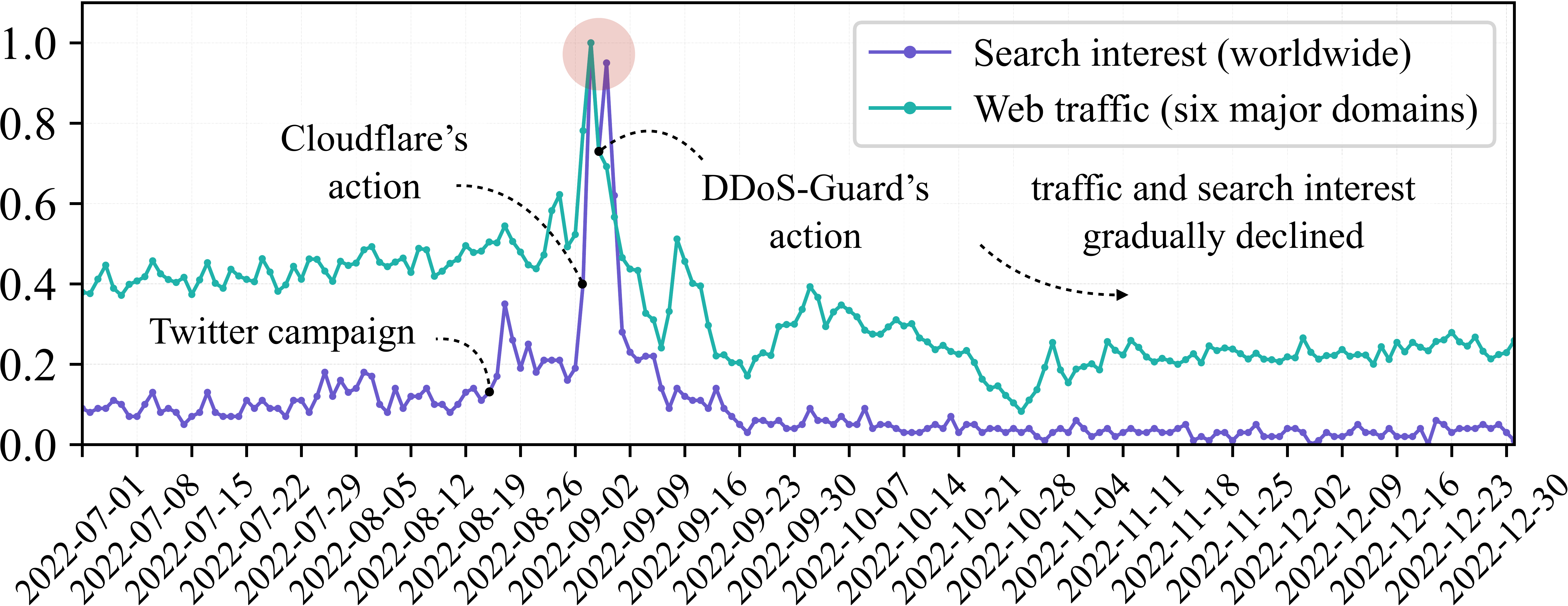}
    \caption{Normalised levels of global search and web traffic to \kiwifarms. The red bubble indicates the Streisand effect.}
    \label{fig:search-trends-and-traffic}
\end{figure}
\begin{figure*}[t]
    \centering
    \includegraphics[width=\textwidth]{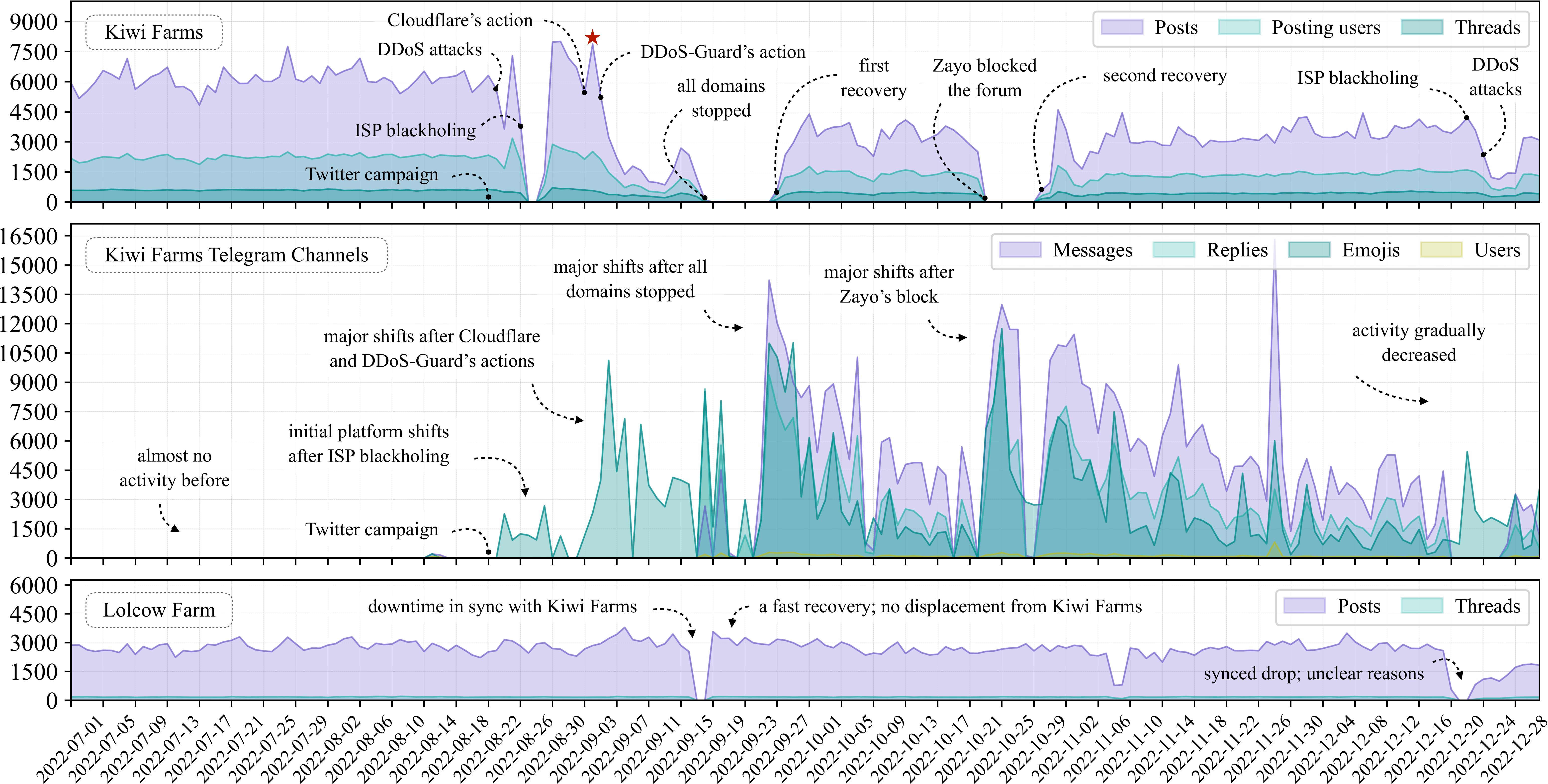}
    \caption{Number of daily posts, threads, and active users on \kiwifarms, its Telegram channels, and its primary competitor \lolcowfarms, as well as major disruptions and displacement between platforms. The red star indicates the Streisand effect.}
    \label{fig:platform-displacement}
\end{figure*}

\section{The Impact on Forum Activity and Traffic} \label{sec:forum-activity-and-displacement}
\noindent On 3 September 2022, \cloudflare discontinued its DDoS prevention service, which attracted major publicity. This intervention led to a sudden and significant increase in global search interest about \kiwifarms with a seven-fold spike, along with the web traffic to the six major domains doubling on 4 September 2022 (see Figure~\ref{fig:search-trends-and-traffic}). This phenomenon, known as the Streisand effect, might be caused by people's curiosity about what happened to the platform, which is relatively rare but mainly seen with `freedom of speech' issues~\cite{hutchings2016taking}. It suggests that attempts at censorship may eventually end up being counterproductive~\cite{jansen2015streisand}: disruptive effort aiming to reduce user interactions instead led to the unintended consequences of increased attention, despite lasting for only a few days before declining sharply. 

We examine in detail the impacts of the disruption and the forum recovery on \kiwifarms within 6 months from July to December 2022. This timeframe provides a sufficient understanding, as the campaign was mostly over and the forum was growing stably before the disruption. To assess the impacts, we separate the observed data points into the post-disruption period (first group) and pre-disruption period (second group), split by 3 September 2022. We then use the Mann–Whitney U test (as the samples are not paired and the data does not follow a normal distribution) to compare the difference between mean ranks of the two populations. The effect size -- indicating the magnitude of the observed difference -- is assessed by Cliff's Delta, ranging [-1, 1].

\subsection{The Impact of Major Disruptions}
\noindent While some DDoS attacks were large enough to shut the forum down, their impact was temporary. For example, the DDoS attack on 23 August 2022 -- which was probably associated with the Twitter campaign the previous day -- led to a drop of roughly 35\% in posting volume, yet the forum activity recovered the next day to a slightly higher level (see the first graph of Figure~\ref{fig:platform-displacement}). The DDoS attack during Christmas 2022 was also short-lived. The ISP blackholing on 26 August 2022 was more critical, silencing the forum for two consecutive days, yet it again recovered quickly.

The most significant, long-lasting impact was caused by the substantial industry disruption that we analyse in this paper. While forum activity immediately dropped by around 20\% after \cloudflare's action on 3 September 2022, the forum was still online at \url{kiwifarms.ru}, hosting the same content. Activity did not degrade significantly until \ddosguard's action on 5 September 2022, which took down the Russian domain. By 18 September 2022, all domains were unavailable, including .onion (presumably their hosting was identified); forum activity dropped to zero and stayed there for a week. The operator managed to get the forum back online for the first time on 27 September 2022, after which it ran stably on both the dark web and clear web for roughly one month until Zayo -- a Tier-1 ISP -- blocked it on 23 October 2022. This led to another silent week before the forum eventually recovered a second time on 30 October 2022. It has been stable since then without serious downtime except for the ISP blackholing on 22 December 2022 which led to a 70\% drop in activity. 

In general, although the forum is now back online stably, hosted on 1776 Solutions -- a company also founded by the forum's owner -- it has (at the moment) failed to bounce back to the pre-disruption level, with the number of active users and posting volume roughly halved. The concerted effort we analysed was much more effective than previous DDoS attacks, yet still could not silence the forum for long.

\begin{figure*}[t]
    \centering
    \includegraphics[width=\textwidth]{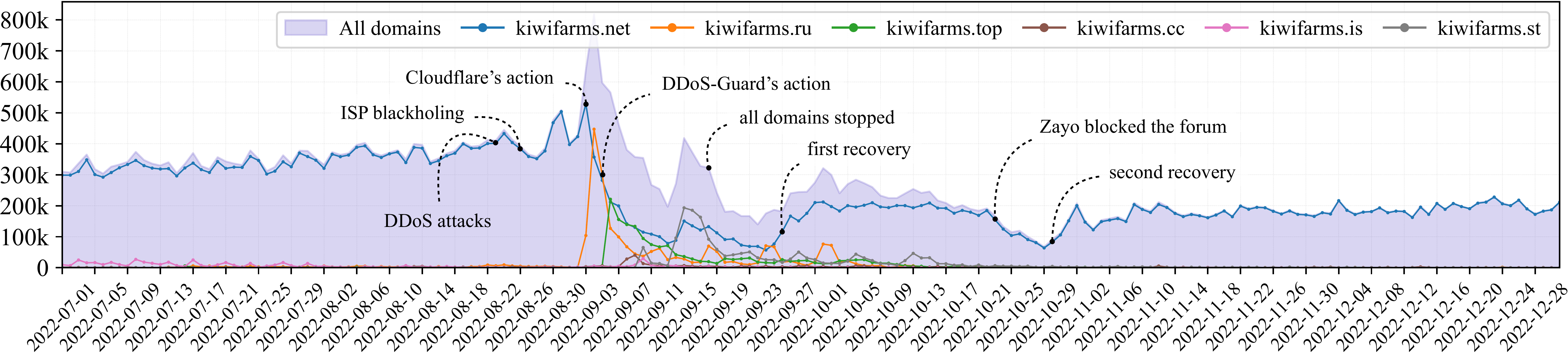}
    \caption{Number of daily estimated visits to \kiwifarms and the fragmentation to previously abandoned domains. We see non-zero traffic to the primary domain when the forum was down, presumably Similarweb counted unsuccessful attempts.}
    \label{fig:traffic-statistics}
\end{figure*}
\subsection{Platform Displacement} 
\noindent The natural behaviour of online communities when their usual gathering place becomes inaccessible is to seek alternative places or channels to continue their discussions. The second graph in Figure~\ref{fig:platform-displacement} illustrates an initial shift of forum activity to Telegram that occurred on 27 August 2022, right after the ISP blackholing. This was accompanied by thousands of emoji reactions on the admin's announcements since commenting was not allowed at that time. Community reactions (e.g., replies, emojis) seem to have been consistent with the overall Telegram posting activity, which increased rapidly afterwards and even occasionally surpassed the forum's activity, especially after the publicity given to the \cloudflare and \ddosguard actions. At some point, for instance in early October and November 2022, the total number of messages on \kiwifarms and its Telegram channels significantly exceeded the pre-disruption posting volume on the forum. However, significant displacements only occurred when all domains were completely inaccessible on 18 September 2022, and again when Zayo blocked the second incarnation of the forum on 22 October 2022. The shift to the Telegram channels appears to be rapid yet rather temporary: \kiwifarms users quickly returned to the primary forum when it became available, while discussion activity on the Telegram channels gradually declined. 

There was no significant shift in activity from the forum to its primary competitor \lolcowfarms (see the third graph of Figure~\ref{fig:platform-displacement}), however, there was an increase in posting on \lolcowfarms about the incident, indicating a minor change of discussion topic (see more in~\S\ref{subsec:forum-members}). It is unclear if these posting users migrated from \kiwifarms, as \lolcowfarms do not use handles, making user counts unavailable. \lolcowfarms also experienced downtime on 17 and 18 September 2022 (the same day as \kiwifarms) yet we have no reliable evidence to draw any convincing explanation. Another drop occurred around Christmas 2022 in sync with \kiwifarms, perhaps because of the holiday. The activity of \lolcowfarms returned to its previous level quickly after these drops, suggesting that the campaign did not significantly impact \lolcowfarms or drive content between the rival ecosystems; the displacement we observed on \kiwifarms was mostly `internal' within its own ecosystem, rather than an `external' shift to other forums. While the disruption impact on \kiwifarms and its Telegram channels is highly significant with a very large effect size, it is \textit{not} significant for \lolcowfarms~(the effect size is small), see Table~\ref{tab:statistical-significance-tests-forum-activity}.

\begin{table}[t]
\centering
\small
\caption{The significance of the disruption in daily activities of \kiwifarms, its Telegram channels, and \lolcowfarms.}
\setlength{\tabcolsep}{.45em}
\begin{tabular}{lrrrr}
\toprule
\small{Platforms} & \small{Variables} & \small{Mann–Whitney U tests} & \small{$\delta$}\\
\midrule
\multirow{3}{*}{\kiwifarmssmall} & \# posts & $U = \KFnKiwiFarmsPostsSignificanceTestStat, \KFnKiwiFarmsPostsSignificanceTestPValue$ & \KFnKiwiFarmsPostsEffectSize\\
& \# threads & $U = \KFnKiwiFarmsThreadsSignificanceTestStat, \KFnKiwiFarmsThreadsSignificanceTestPValue$ & \KFnKiwiFarmsThreadsEffectSize\\
& \# users & $U = \KFnKiwiFarmsUsersSignificanceTestStat, \KFnKiwiFarmsUsersSignificanceTestPValue$ & \KFnKiwiFarmsUsersEffectSize\\
\arrayrulecolor{black!20}
\midrule
\multirow{4}{*}{\shortstack[l]{\kiwifarmssmall\\Telegram}} & \# messages & $U = \KFnTelegramMessagesSignificanceTestStat, \KFnTelegramMessagesSignificanceTestPValue$ & \KFnTelegramMessagesEffectSize\\
& \# replies & $U = \KFnTelegramRepliesSignificanceTestStat, \KFnTelegramRepliesSignificanceTestPValue$ & \KFnTelegramRepliesEffectSize\\
& \# emojis & $U = \KFnTelegramEmojisSignificanceTestStat, \KFnTelegramEmojisSignificanceTestPValue$ & \KFnTelegramEmojisEffectSize\\
& \# users & $U = \KFnTelegramUsersSignificanceTestStat, \KFnTelegramUsersSignificanceTestPValue$ & \KFnTelegramUsersEffectSize\\
\arrayrulecolor{black!20}
\midrule
\multirow{2}{*}{\lolcowfarmsmall} & \# posts & $U = \KFnLolcowFarmPostsSignificanceTestStat, \KFnLolcowFarmPostsSignificanceTestPValue$ & \KFnLolcowFarmPostsEffectSize\\
& \# threads & $U = \KFnLolcowFarmThreadsSignificanceTestStat, \KFnLolcowFarmThreadsSignificanceTestPValue$ & \KFnLolcowFarmThreadsEffectSize\\
\arrayrulecolor{black}
\midrule
\end{tabular}
\label{tab:statistical-significance-tests-forum-activity}
\vspace{-2mm}\end{table}
\begin{table}[t]
\centering
\small
\caption{The significance of the disruption in daily traffic to each individual domain of \kiwifarms, and to all domains.}
\setlength{\tabcolsep}{.65em}
\begin{tabular}{lrrrr}
\toprule
\small{Platforms} & \small{Variables} & \small{Mann–Whitney U tests} & \small{$\delta$}\\
\midrule
kiwifarms.net & \# visits & $U = \KFkiwifarmsnetSignificanceTestStat, \KFkiwifarmsnetSignificanceTestPValue$ & \KFkiwifarmsnetEffectSize\\
kiwifarms.ru & \# visits & $U = \KFkiwifarmsruSignificanceTestStat, \KFkiwifarmsruSignificanceTestPValue$ & \KFkiwifarmsruEffectSize\\
kiwifarms.top & \# visits & $U = \KFkiwifarmstopSignificanceTestStat, \KFkiwifarmstopSignificanceTestPValue$ & \KFkiwifarmstopEffectSize\\
kiwifarms.cc & \# visits & $U = \KFkiwifarmsccSignificanceTestStat, \KFkiwifarmsccSignificanceTestPValue$ & \KFkiwifarmsccEffectSize\\
kiwifarms.is & \# visits & $U = \KFkiwifarmsisSignificanceTestStat, \KFkiwifarmsisSignificanceTestPValue$ & \KFkiwifarmsisEffectSize\\
kiwifarms.st & \# visits & $U = \KFkiwifarmsstSignificanceTestStat, \KFkiwifarmsstSignificanceTestPValue$ & \KFkiwifarmsstEffectSize\\
\arrayrulecolor{black!20}
\midrule
All domains & \# visits & $U = \KFkiwifarmsAllDomainsSignificanceTestStat, \KFkiwifarmsAllDomainsSignificanceTestPValue$ & \KFkiwifarmsAllDomainsEffectSize\\
\arrayrulecolor{black}
\midrule
\end{tabular}
\label{tab:statistical-significance-tests-web-traffic}
\vspace{-2mm}\end{table}
\subsection{Traffic Fragmentation} \label{subsec:traffic-fragmentation}
\noindent Before \cloudflare's action, traffic towards \kiwifarms (measured by Similarweb)  was relatively steady, mostly occupied by the primary domain. However, we see the Streisand effect with an immediate peak in traffic of around 50\% more visits and 85\% more visitors once the site was disrupted. The publicity given by the takedown presumably boosted awareness and attracted people to visit both the primary and alternative domains. Traffic to the primary domain was then significantly fragmented to other previously abandoned domains, resulting in the \url{kiwifarms.net} accounting for less than 50\% one day after \cloudflare's intervention, as shown in Figure~\ref{fig:traffic-statistics}. 

Following the unavailability of \url{kiwifarms.net}, most traffic was directed to \url{kiwifarms.ru}, which was under DDoS Guard's protection (accounting for around 60\% total traffic on 4 September 2022). The \ddosguard's action on 5 September 2022 reduced traffic towards \url{kiwifarms.ru} sharply, while traffic towards \url{kiwifarms.top} peaked. The suspension of \url{kiwifarms.top} on the following day led to increased traffic towards \url{kiwifarms.cc} (a Pleroma decentralised web instance), but it only lasted for a couple of days before traffic shifted again to \url{kiwifarms.is}. The seizure of \url{kiwifarms.is} later led to the traffic shifting to \url{kiwifarms.st}, but it was also short-lived. 

The forum recovery on 27 September 2022 gradually directed almost all traffic back to the primary domain, and by 22 October 2022, \url{kiwifarms.net} mostly accounted for all traffic, albeit at about half the volume. This effect is highly consistent with what has been found in our forum data, indicating a reliable pattern. Overall, our evidence suggests a clear traffic fragmentation across different domains of \kiwifarms, in which people attempted to visit surviving domains when one was disrupted. While the observed fragmentation is clear, the impact on two domains is not significant enough when assessing the period as a whole. However, it is highly significant on the total traffic, notably the substantial drops of the primary domain \url{kiwifarms.net} (see Table~\ref{tab:statistical-significance-tests-web-traffic}).

\section{The Impacts on Relevant Stakeholders} \label{sec:relevant-stakeholders}
\noindent We have looked at the impacts of the disruption on \kiwifarms itself. This section examines the effects on relevant stakeholders, including the harassed victim, the community leading the campaign, the industry, the forum operators, and active forum users who posted at least once. As our ethics approval does not allow the study of individuals, all measurements are conducted collectively on subsets of users. Besides quantitative evidence, we also qualitatively look at statements made by tech firms about the incident.

\begin{figure}[t]
    \centering
    \includegraphics[width=0.48\textwidth]{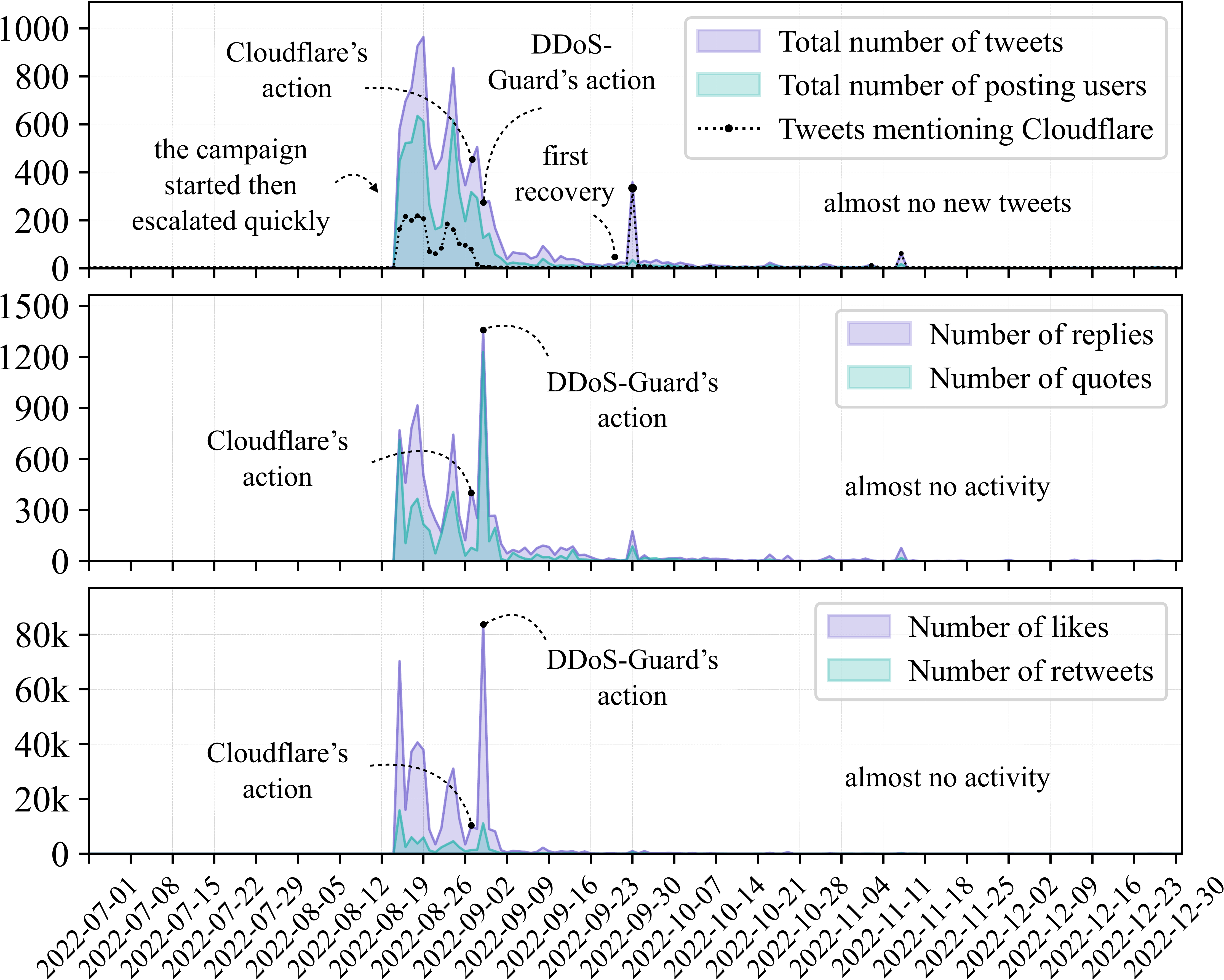}
    \caption{Number of daily tweets and reactions made by the community about the campaign. Figure scales are different.}
    \label{fig:twitter-activity}
\end{figure}
\subsection{The Community that Started the Campaign}
\noindent There were \KFnPostingTweeterUsers~users in the online community involved in starting the campaign. Of these, \KFnTweeterKeyActors~users (\KFnTweeterKeyActorsProps\%)~were responsible for around 80\% of tweets. There was a sharp increase in tweets and reactions at the beginning (see Figure~\ref{fig:twitter-activity}). The first peak was on 25 August 2022 with nearly 900 tweets by around 600 users. However, this dropped rapidly to less than 100 per day after a few weeks when \cloudflare and \ddosguard took action, and almost to zero two weeks later. The number of tweets specifically mentioning \cloudflare (such as their official account, as well as those for jobs, help, and developers) was around 200 in the beginning but decreased over time, and dropped to zero after they took action. This lasted for roughly one month until after the forum recovered: we see around 400 tweets mentioning \cloudflare, twice the previous peak, and accounting for almost all such tweets that day. However, these tweets appeared to be primarily associated with another campaign counted by the hashtag \#stopdoghate, suggesting a short-lived outlier instead of a genuine peak.

The trans activist who launched the campaign was engaged at the beginning but then became much less active in posting new tweets, although she still replied to people. Her posting volume was, however, trivial compared to the overall numbers: she made only four tweets on the day the campaign started, the number then dropped quickly to only one on 4 September 2022 after \cloudflare took action, and zero thereafter. It suggests that although she sparked the campaign, she might not be the primary maintainer. 

We see no notable peak of tweets after the forum was completely shut down, suggesting a clear loss of interest in pursuing the campaign, both from people posting tweets and people reacting to tweets. The community seemed to get bored quickly after a few weeks when they appeared to have gotten what they wanted -- `\textit{Kiwi Farms is dead, and I am moving on to the next campaign}', tweeted the activist. 

\subsection{The Industry Responses}
\noindent Unlike measurement of forum activity, there is no such quantitative data to cover the impact on industry actors, so we switch to qualitative analysis of public statements made by those who directly attempted to terminate the forum. We first compile a list of involved tech firms seen from the takedown incidents, then look at their websites, news, and blogs to spot their official statements if available. We repeated the search regularly, and the final list consists of four firms: \cloudflare, \ddosguard, \diamwall, and \harica. We then took a deductive approach to understand (1) their hosting policy, (2) their perspective on \kiwifarms, and (3) their reactions to the community pressure.

\cloudflare stated their abuse policies on 31 August 2022 without directly mentioning the Twitter campaign~\cite{cloudflareabusepolicies}. In summary, the firm offers traffic proxy and DDoS protection to lots of (mostly non-paid) sites regardless of the content hosted, including \kiwifarms. The firm maintains that abusive content alone is not an issue, and the forum -- while immoral -- still deserves the same protection as other customers, as long as it does not violate US law. Although \cloudflare are entitled to refuse business from \kiwifarms, they initially took the view that doing so because of its content would create a bad precedent, leading to unintended consequences on content regulation and making things harder for \cloudflare. This could affect the whole Internet, as \cloudflare handles a large proportion of network traffic. They did not want to get involved in policing online content, but if they had to do it they would rather do so in response to a court order instead of popular opinion. The firm previously had dropped the neo-Nazi website \dailystormer~\cite{cloudflaretakedowndailystormer} and the extremist board \eightchan~\cite{cloudflaretakedown8chan} because of their links with terrorist attacks and mass murders, and a false claim about \cloudflare's secret support. They also claimed that dropping service for \kiwifarms would not remove the hate content, but only slow it down for a while.

Nevertheless, \cloudflare did a U-turn a few days later on 3 September 2022, announcing that they would terminate service for \kiwifarms~\cite{cloudflarebloackskiwifarms}. They explained that the escalation of the pressure campaign led to users being more aggressive, which might lead to crime. They reached out to law enforcement in multiple jurisdictions regarding potential criminal acts, but as the legal process was too slow compared to the escalating threat, they made the decision alone~\cite{cloudflarebloackskiwifarms}. They still claimed that following a legal process would be the correct policy, and denied that the decision was the direct result of community pressure. \cloudflare's action also inadvertently led to the termination of a neo-Nazi group in New Zealand, as it was hosted by the same company as the forum~\cite{keffalsCFdeplatformedneonazigroup}.

\ddosguard's statements about the incident told a similar story~\cite{kiwifarmsblockedbyddosguard}. Although they can restrict access to their customers if they violate the acceptable use policy, content moderation is not their duty (except under a court order) so they do not need to determine whether every site they protect violates the law. \diamwall took the same line; they claimed that they are not responsible, and are unable to moderate content hosted on websites~\cite{kiwifarmsblockedbydiamwall}. They also maintained that terminating services in response to public pressure is not good policy, but the case of \kiwifarms was exceptional due to its `revolting' content. They also noted that their actions could only delay things but not fix the root cause, as the forum could find another provider. \diamwall's statement was removed afterwards without any clear explanations; it is now only accessible through online archives. 

Unlike the three firms above, \harica~-- a Greek Authority providing certificates for .onion sites -- took a different line. They confirmed their support for freedom of speech and stated that they will not censor any website, but are obligated to investigate complaints about websites violating the law, the Certificate Policy (CP) and the Certificate Practice Statement (CPS). After a review process, on 15 May 2023 they announced they would revoke the .onion certificates issued to \kiwifarms due to concerns about harassment connecting to suicides. After setting a 3-day timeframe for \kiwifarms to seek a new authority, their support team were targeted by various threats and harassment, but based on one `kind and polite' message highlighting that \harica is one of the only two authorities issuing .onion certificates, they postponed the decision the day after and waited for further law enforcement investigation as \kiwifarms has very limited alternatives to protect their site~\cite{haricaannouncement}.

Whether blocking \kiwifarms or not, it is understandable that infrastructure and certificate providers may not want to get involved in content regulation the way Facebook and Google have to, as moderation is complex, challenging, contentious and expensive~\cite{bromell2022challenges}.

\subsection{The Forum Operators} \label{subsec:forum-operators}
\begin{figure}[t]
    \centering
    \includegraphics[width=0.48\textwidth]{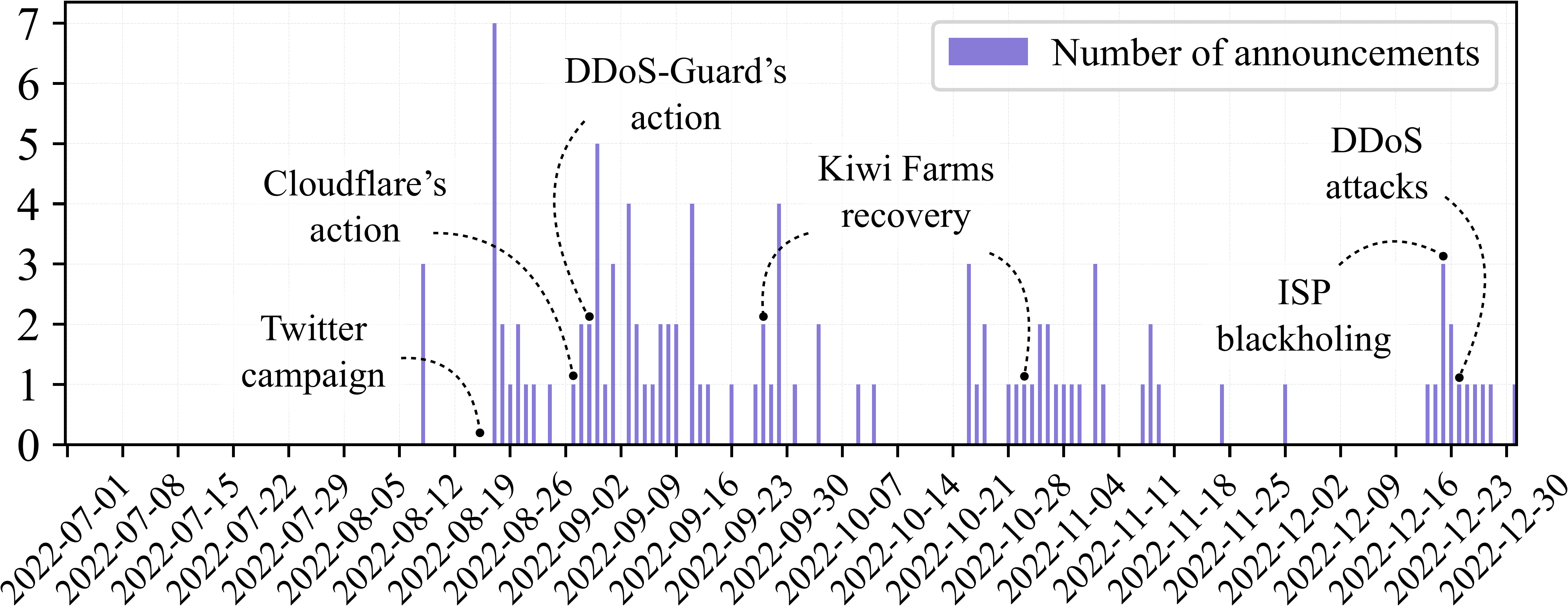}
    \caption{Number of public announcements posted daily by the forum operators since the Telegram channel was created.}
    \label{fig:operators-announcements}
\end{figure}
\noindent The disruption of \kiwifarms led to a cat-and-mouse game where tech firms tried to shut it down by various means while the forum operators tried to get it back up. We extract messages of the forum operators from a Telegram channel activated after the Twitter campaign, where they posted \KFnOperatorsAnnouncements~announcements during the period, mostly about when and where the forum was back, the ongoing issues (e.g., DDoS attacks, industry blocks), and their plans to fix.

The admins were very active, for example, sending seven consecutive messages on 23 August 2022 that mostly concerned the large DDoS attack on that day, see Figure~\ref{fig:operators-announcements}. The second peak was on 6 September 2022 after \cloudflare and \ddosguard's withdrawal of service, mostly about forum availability. The number of announcements then gradually decreased, especially after the second recovery, with many days having no messages. A DDoS attack hitting the forum during Christmas 2022 caught the admins' attention for a while. Their activity was inversely correlated with the forum's stability; they were less active when the site was up and running stably or when there were no new incidents, for example, many announcements were posted in September, late October, and late December 2022, when the forum was under DDoS attacks and disruptions as shown in Figure~\ref{fig:platform-displacement}.

We took a deductive approach based on the extracted announcements to comprehend the effort made by the forum operator to restore service. \kiwifarms needed DDoS protection to hide its original IP address and evade cyberattacks, so the operators first switched their third-party DDoS protection to \ddosguard, then \diamwall, yet these firms also resigned their business. They then attempted to build an anti-bot mechanism themselves based on HAProxy -- an open-source software to stop bots, spam, and DDoS using proof-of-work~\cite{haproxy} -- and claimed to be resilient to thousands of simultaneous connections. They also changed hosting providers to VanwaTech and eventually their own firm 1776 Solutions, and attempted to route their traffic through other ISPs. They were actively maintaining infrastructure, fixing bugs, and giving instructions to users to deal with their passwords when the forum experienced a breach. The operators' effort seemed to be competent and consistent.

\subsection{The Forum Members} \label{subsec:forum-members}
\noindent People sharing the same passion naturally coalesce into communities, in which some key actors may play a crucial role in influencing the ecosystem~\cite{hughes2019playing,van2021heterogeneity,vu2020turning}. We separate the pre-disruption and post-disruption by 3 September 2022, when \cloudflare took action. \kiwifarms activity is highly skewed, with around 80\%\footnote{We make use of the 80/20 rule -- the Pareto principle~\cite{sanders1987pareto}.} of pre-disruption posts made by \KFnBigFishProps\%~most active users (\KFnBigFish), while the remaining 20\% posts were made by the \KFnSmallFishProps\%~less active (\KFnSmallFish). There was around a 30\% drop in the number of users after the disruption, as seen in Figure~\ref{fig:platform-displacement}.

There were \KFnNewUsernamesIncludingReturningFish~new usernames after the disruption, which could be either newcomers or old members creating new accounts after losing access to old ones. Multi-platform users tend to pick similar pseudonyms on different platforms~\cite{goga2013exploiting}; we believe returning users are also likely do that to preserve their reputation, so can be detected if their usernames are very similar and rare enough, although common handles are often picked up by multiple individuals~\cite{liu2013s}. While the similarity of two usernames can be determined by the Levenshtein distance, we use a n-gram model trained by the Reuters corpus~\cite{russell2002reuters} to estimate the rarity of usernames, considering one is rare if the highest probability observed is not greater than 1\%. We found 5.31\% such users among \KFnNewUsernamesIncludingReturningFish~new pseudonyms: \KFnReturningBigFish~returning core actors (\KFnReturningBigFishProps\% of core users), \KFnReturningSmallFish~returning casual actors (\KFnReturningSmallFishProps\% of casual users), while the rest \KFnNewFish~are newcomers. The estimation may overlook all-new usernames, yet we believe this number is relatively small as a mass password reset was mandated after the breach instead of account replacement.

We analyse the behaviour of those active after the disruption, namely the `core survivors' (\KFnSurvivingBigFish, returning actors included), `casual survivors' (\KFnSurvivingSmallFish, returning actors included), and `newcomers' (\KFnNewFish). Around half of key users (\KFnSurvivingBigFishProps\%) remained engaged, while only \KFnSurvivingSmallFishProps\% of casual users stayed (\KFnLeavingSmallFishProps\% has left). On average, before the disruption, each `core survivor' posted \KFavgPostsPreDisruptionCoreSurvivorsHigherThanCasualSurvivors~times more than each `casual survivor' (\KFavgPostsPreDisruptionCoreSurvivors~vs \KFavgPostsPreDisruptionCasualSurvivors~posts), while their active period -- between their first post and last post -- was around \KFavgLifetimePreDisruptionCoreSurvivorsHigherThanCasualSurvivors~times longer (\KFavgLifetimePreDisruptionCoreSurvivors~vs \KFavgLifetimePreDisruptionCasualSurvivors~days).

\begin{figure}[t]
    \centering
    \includegraphics[width=0.48\textwidth]{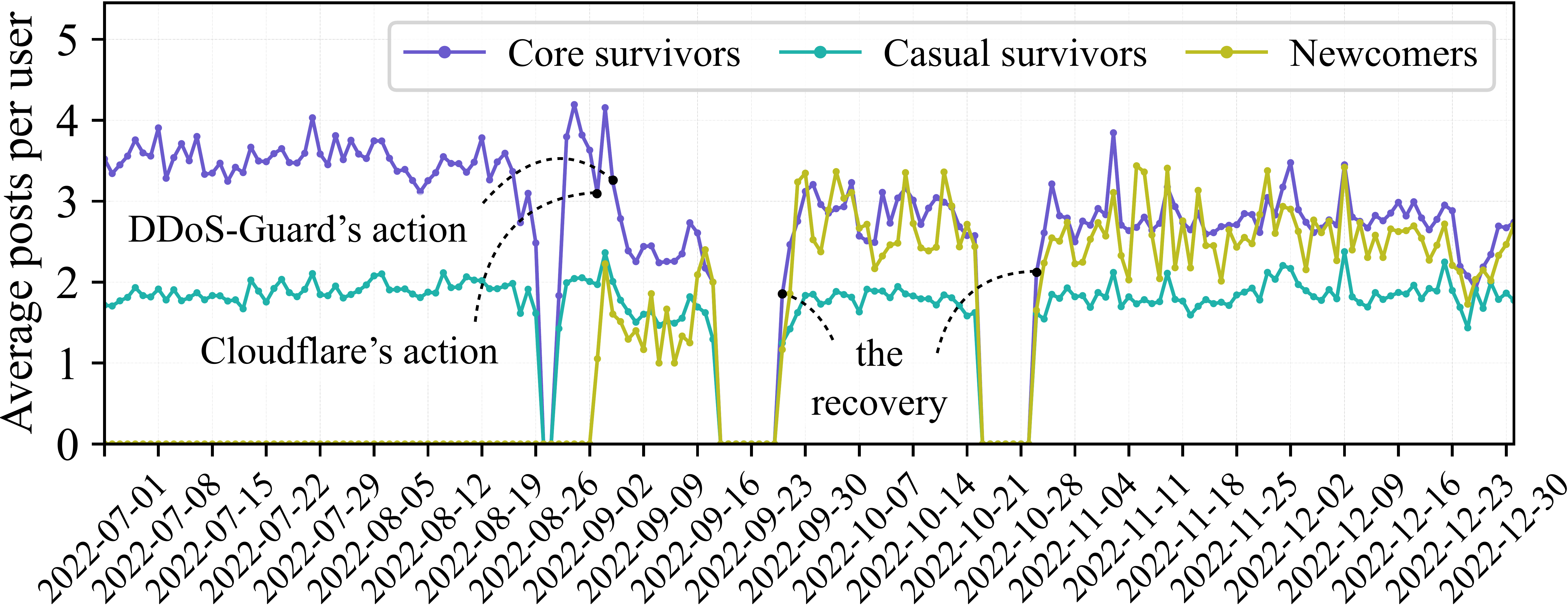}
    \caption{Number of average posts per day of survivors and newcomers, who posted at least once after the disruption.}
    \label{fig:survivors-posting-volume}
\end{figure}
\begin{figure}[t]
    \centering
    \includegraphics[width=0.48\textwidth]{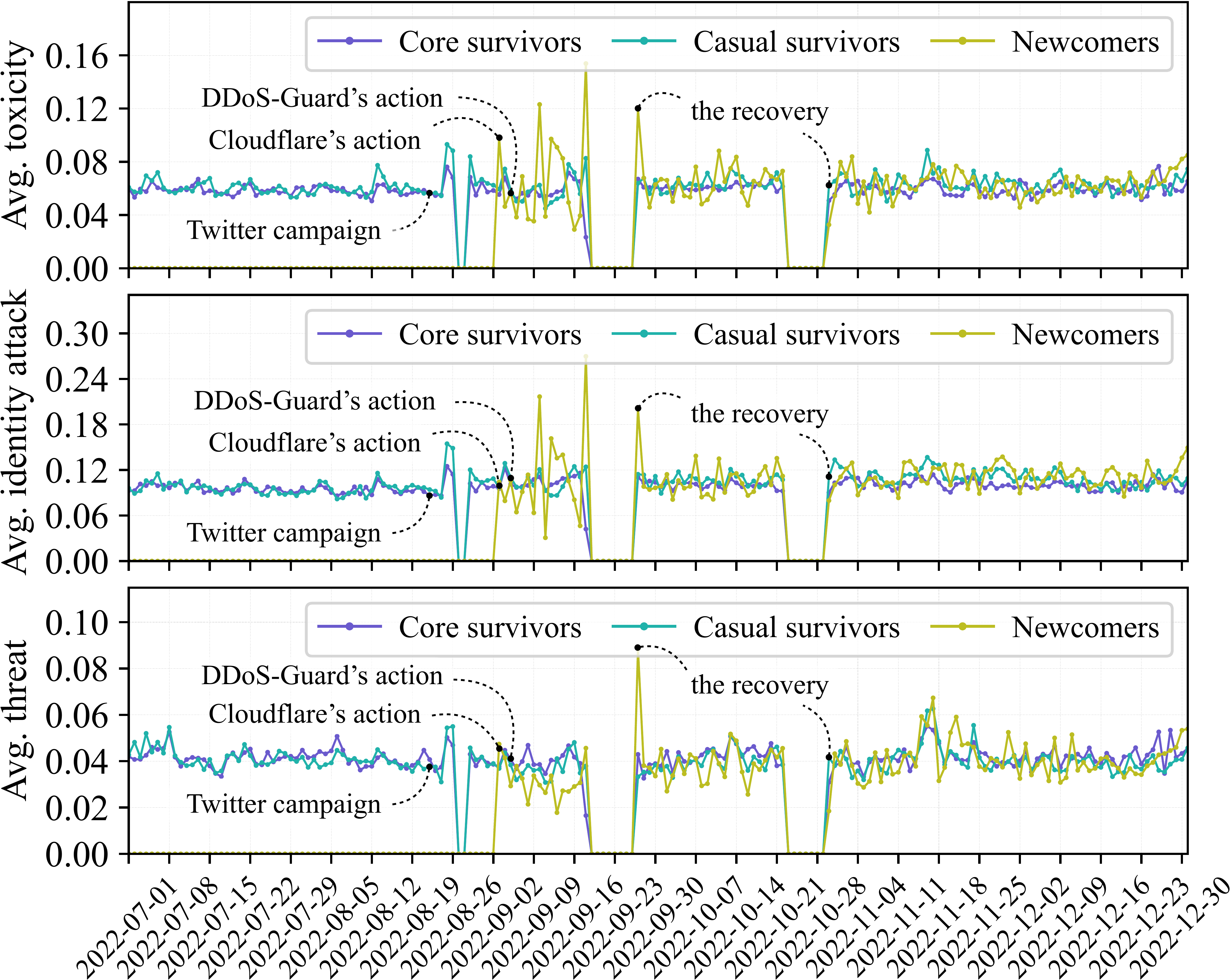}
    \caption{Average toxicity, identity attack, and threat levels of posts made by survivors and newcomers after the event.}
    \label{fig:survivors-toxicity}
\end{figure}
\subsubsection*{Posting Activity} Before the takedown, each core survivor made about 3.5 posts per day on average, while it was around 3 afterwards -- see Figure~\ref{fig:survivors-posting-volume}. The activity of the other survivors appears consistent with the pre-disruption period; their average posts were at around 2 per day before the incident and almost unchanged afterwards. These figures suggest that the decreasing posting volume seen in Figure~\ref{fig:platform-displacement} was mainly due to users leaving the forum, instead of surviving ones largely losing interest -- they engaged back quickly after the forum recovered. Newcomers posted slightly less than casual survivors before the forum was completely down on 18 September 2022 (less than 2 posts per day), yet their average posting volume then increased quickly. This suggests that the disruption, besides removing a very large proportion of old casual users, drew in many new users who then became roughly as active as the core survivors.

\subsubsection*{Toxicity Levels} We further examine the toxicity of posts made by the surviving actors and newcomers, before and after the disruption. Figure~\ref{fig:survivors-toxicity} shows the average levels of \textit{toxicity}, \textit{identity attack} and \textit{threat} of core survivors, casual survivors, and newcomers by days. In general, the \textit{toxicity}, \textit{identity attack}, and \textit{threat} scores were rather low as most postings are non-toxic (despite some having very high scores). There were small changes in the average scores of surviving actors, notably the peaks occurred 2 days after the campaign sparked on Twitter, with the average scores increasing significantly to around 30--50\%, especially \textit{toxicity} and \textit{identity attack}. However, these dropped quickly a couple of days after and retreated to normal levels. 

Newcomers, on the other hand, expressed a significant increase of \textit{toxicity} and \textit{identity attack} during the first two weeks after the disruption took place (about 2--2.5 times higher), largely surpassing surviving actors. Their scores for \textit{threat} did not increase at that time but largely peaked after the forum first recovered on 27 September 2022, with around 2 times higher. These activities suggest that while the surviving members were becoming more toxic when their community was under attack, new users became much more toxic for a few weeks after they engaged in the discussion before declining gradually to the same levels as old users. This is in line with the recent finding that users moving to other platforms can become more toxic than before~\cite{ali2021understanding}.

\subsubsection*{Social Interactions} To measure how these survivors interact with each other, we build a social interaction network among \kiwifarms members over time. We consider each active user as a node, with an edge between two users if they posted in the same thread (weighted by the number of such interactions)~\cite{pete2020social}. We then explore changes of that shared-interest engagement in the network structure with a focus on Degree Centrality, which indicates how well-connected a user is over the entire network~\cite{newman2018networks}. In a healthy community, such engagement should grow steadily.

\begin{figure}[t]
    \centering
    \includegraphics[width=0.48\textwidth]{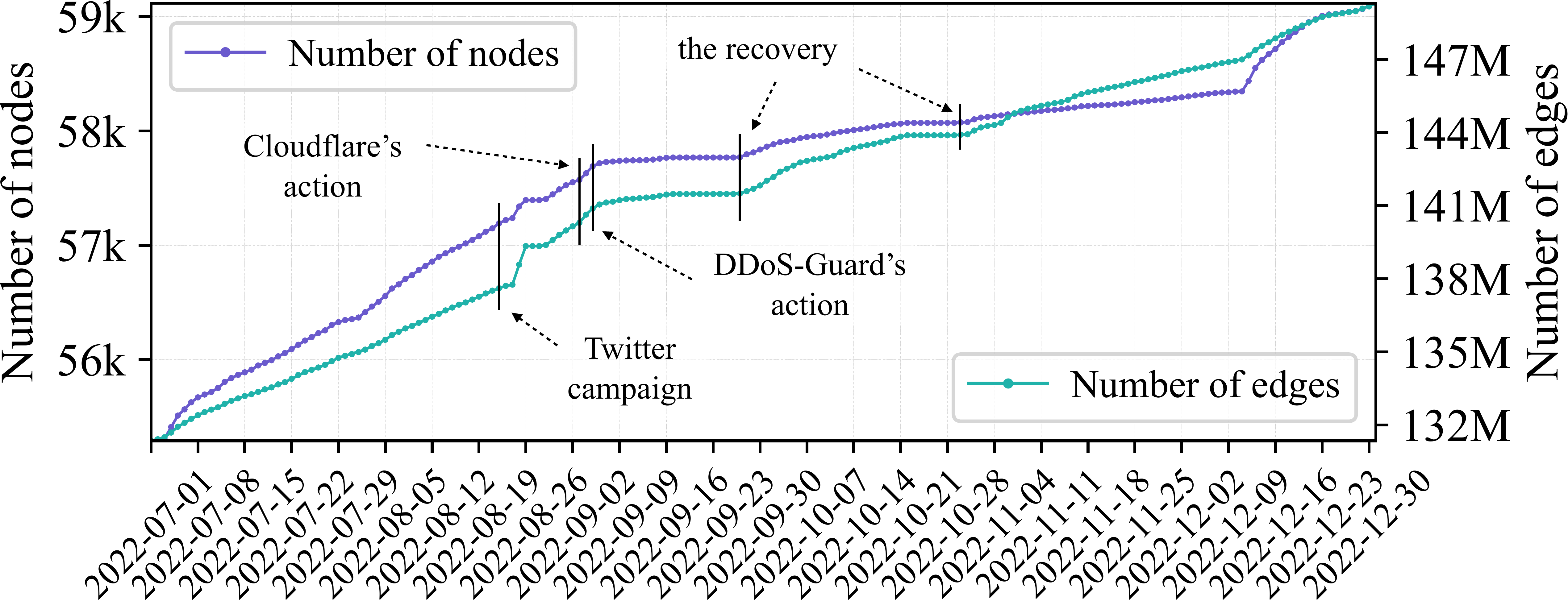}
    \caption{Number of nodes and edges in the social interaction network made by \kiwifarms members over time.}
    \label{fig:survivors-graph-properties}
\end{figure}
\begin{table}[t]
    \centering
    \small
    \caption{Number of posts mentioning the two major involved parties during the period, with proportions of the total posts.}
    \setlength{\tabcolsep}{.475em}
    \begin{tabular}{lrrr}
    \toprule
    \small{Platforms} & \thead[r]{\small{Mentioning}\\\kiwifarmssmall} & \thead[r]{\small{Mentioning}\\\small{\cloudflaresmall}} & \thead[r]{\small{Mentioning}\\\small{both parties}} \\
    \midrule
    \kiwifarmssmall & \KFnKFPostsMentioningKF~(\KFnKFPostsMentioningKFProps\%) & \KFnKFPostsMentioningCF~(\KFnKFPostsMentioningCFProps\%) & \KFnKFPostsMentioningBothKFCF~(\KFnKFPostsMentioningBothKFCFProps\%)  \\
    Telegram & \KFnTelePostsMentioningKF~(\KFnTelePostsMentioningKFProps\%) & \KFnTelePostsMentioningCF~(\KFnTelePostsMentioningCFProps\%) & \KFnTelePostsMentioningBothKFCF~(\KFnTelePostsMentioningBothKFCFProps\%) \\
    \lolcowfarmsmall & \KFnLFPostsMentioningKF~(\KFnLFPostsMentioningKFProps\%) & \KFnLFPostsMentioningCF~(\KFnLFPostsMentioningCFProps\%) & \KFnLFPostsMentioningBothKFCF~(\KFnLFPostsMentioningBothKFCFProps\%) \\
    \bottomrule
    \end{tabular}
    \label{tab:discussion-mentioning}
\end{table}
\begin{figure}[t]
    \centering
    \includegraphics[width=0.48\textwidth]{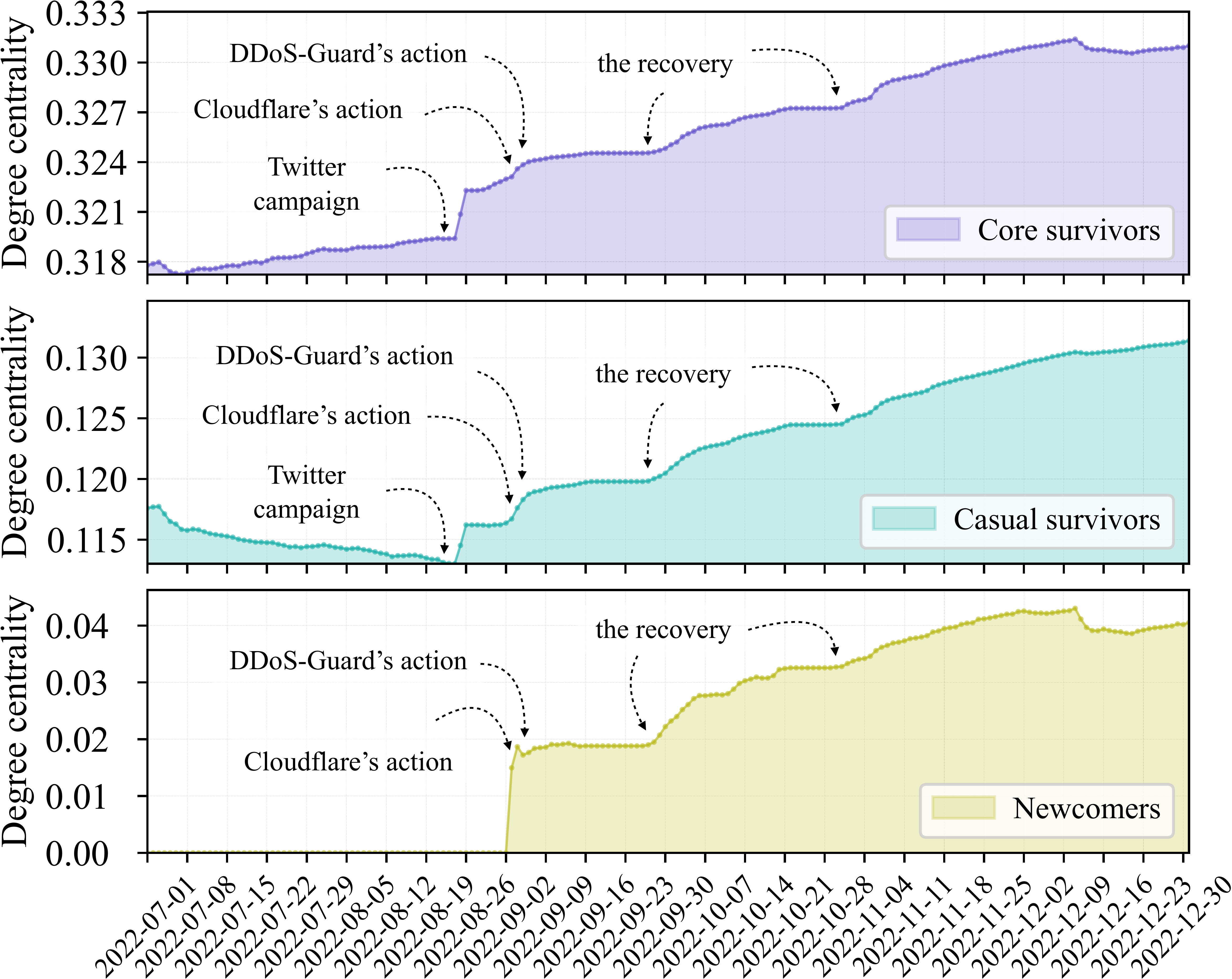}
    \caption{The degree centrality of survivors and newcomers in the network over time. Figures are in different scales.}
    \label{fig:survivors-centrality}
\end{figure}
The network had developed stably before the disruption, with around \KFnNodesAtTheBeginning~nodes and \KFnEdgesAtTheBeginning~edges on 1 July 2022, reaching to around \KFnNodesWhenTwitterCampaign~nodes and \KFnEdgesWhenTwitterCampaign~edges just before the Twitter campaign started (see Figure~\ref{fig:survivors-graph-properties}). There was a rapid increase in both nodes and edges shortly after the Twitter campaign, suggesting that the campaign drew more actors involved in interacting with others. The \cloudflare and \ddosguard actions paused the network for a few weeks, yet it resumed shortly after the forum's recovery. As of 31 December 2022, the network size has reached \KFnNodesAtTheEnd~nodes and \KFnEdgesAtTheEnd~edges. 

Core users are better connected than casual users, see Figure~\ref{fig:survivors-centrality}. The Twitter campaign largely boosted the centrality of both core and casual survivors. Before that, while core survivors were getting more centralised over time, casual survivors were becoming less centralised. But after the campaign on Twitter, the centralisation of both steadily increased. Newcomers came into play quickly afterwards and the forum recovery also made them more centralised.

\subsubsection*{Discussion of the Incident} We examine how users mention the two major involved parties (\kiwifarms and \cloudflare) during the period by extracting posts containing case-insensitive keywords `\textit{kiwifarm}', `\textit{kiwi farm}', `\textit{cloudflare}', and `\textit{cloud flare}' from \kiwifarms, its Telegram channel, and \lolcowfarms. Table~\ref{tab:discussion-mentioning} shows that discussions about the two parties were highly skewed and significantly dependent on the platforms. Telegram users tended to discuss things relevant to \kiwifarms more than \cloudflare (\KFnTelePostsMentioningKFHigherThanCF~times higher), while the ratios were less skewed for \kiwifarms and \lolcowfarms, with \KFnKFPostsMentioningKFHigherThanCF~and \KFnLFPostsMentioningKFHigherThanCF, respectively. These discussions are centralised around a small number of conversations, for example, over 50\% of posts mentioning \cloudflare on \kiwifarms are just from 4 threads.

\begin{figure}[t]
    \centering
    \includegraphics[width=0.48\textwidth]{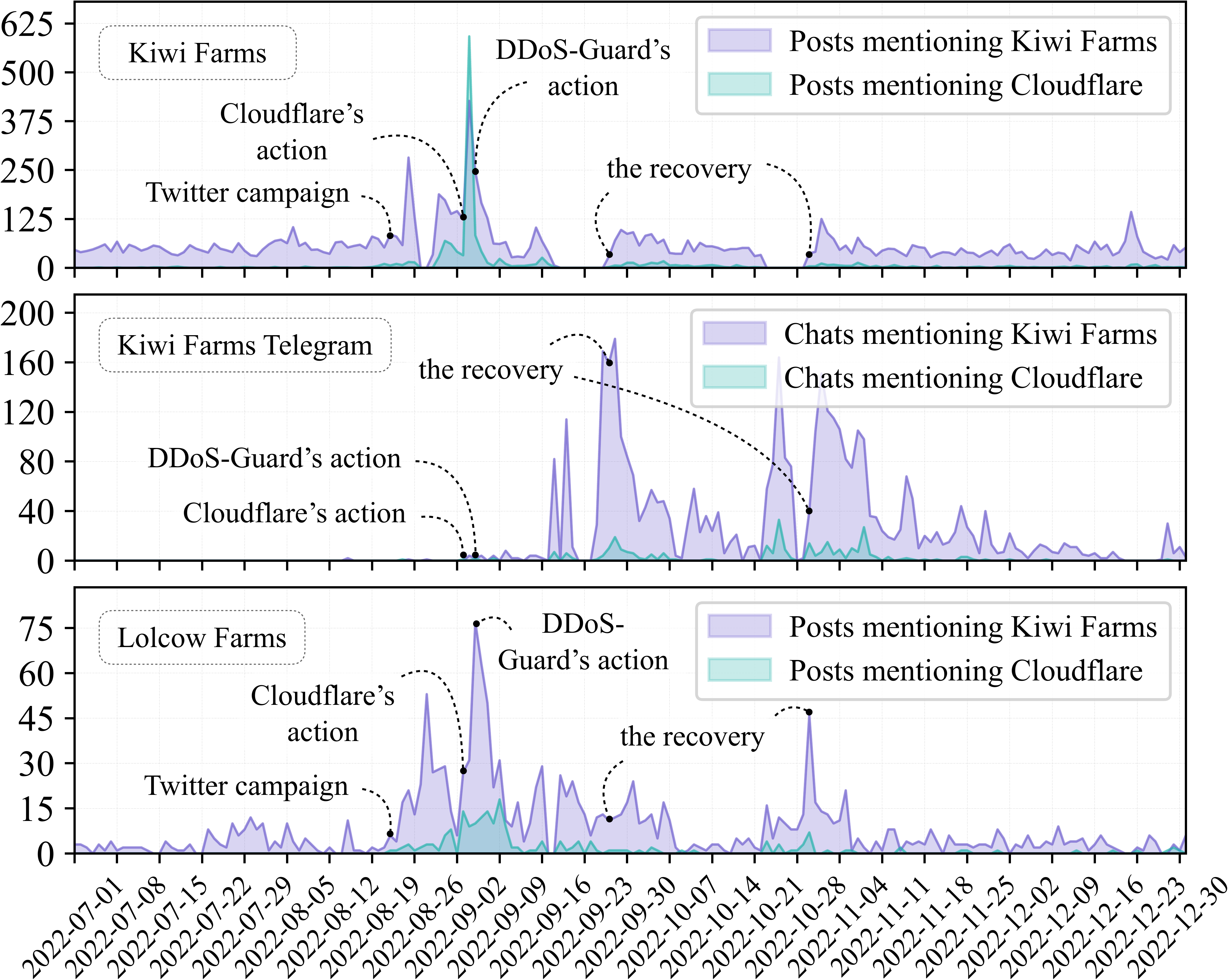}
    \caption{Number of daily posts regarding the disruption on \kiwifarms, its Telegram channels, and \lolcowfarms.}
    \label{fig:discussion-mentioning}
\end{figure}

Although these posts accounted for a trivial contribution to the total posting volume on all three platforms as shown in Figure~\ref{fig:platform-displacement}, most happened after the Twitter campaign, with almost no discussion before. The topic was popular for a short period, as shown in Figure~\ref{fig:discussion-mentioning}. Users on both forums started discussing the incident shortly after the campaign started on 22 August 2022. The topic was energised on both forums after \cloudflare's action on 3 September 2022, peaking on 4 September 2022 on \kiwifarms with over 400 and 600 posts about \kiwifarms and \cloudflare (around 5\% and 7.5\% of all posts on that day), respectively. After \kiwifarms activity was significantly reduced due to \ddosguard's action on 5 September 2022, posts mentioning \kiwifarms and \cloudflare on \lolcowfarms peaked at around 80 and 20, respectively.\footnote{The numbers for \lolcowfarmsmini are typically lower than \kiwifarmsmini as \lolcowfarmsmini is smaller and centred on images instead of text. We do not collect images for safety and ethical reasons, but we believe the trends observed are likely indicative if not reliable.} Telegram activity regarding the incident was a bit different, as comments were only allowed after the forum was completely down; it followed the same trends as overall activity, with a peak of discussion about \kiwifarms happening largely when the forum was inaccessible, as part of the discussion had moved here.

Discussion mentioning \kiwifarms greatly exceeded those mentioning \cloudflare until the day \cloudflare took action (see the first graph in Figure~\ref{fig:discussion-mentioning}). The pattern seen on \lolcowfarms suggests that the attention toward the incident was reflected there, although the peak did not correlate with the overall volume observed in Figure~\ref{fig:platform-displacement} as this contribution is trivial compared to the total. There were almost no posts about \cloudflare after \kiwifarms became completely inaccessible, but there were still around 20 posts about \kiwifarms seen on \lolcowfarms during that week. While nothing changed on \kiwifarms during the second recovery, there was an increase in posts on \lolcowfarms about the incident, presumably as people got the news.

Overall, attention on \kiwifarms, its Telegram channels, and \lolcowfarms was directed to the incident by the Twitter campaign, with posting volume peaking after the industry action. We believe it shows a genuine effect as none of the users there discussed \cloudflare and \kiwifarms before. However, the effect was temporary and almost dropped to the pre-disruption level after the second recovery: they lasted for a few days on \kiwifarms, around one week on \lolcowfarms (partly due to many domains of \kiwifarms being down while \lolcowfarms was still active), and a few weeks on Telegram. Users' interest was fleeting; they largely stopped talking about the incident after a few weeks. 

\section{Tensions, Challenges, and Implications} \label{sec:discussion} 
\noindent The disruption we analysed could be the first time a series of infrastructure firms were involved in a collective effort to terminate a website. While deplatforming can reduce the spread of abusive content and safeguard people's mental and physical safety, and is already routine on social-media platforms like Facebook, doing so without due process raises a number of philosophical, ethical, legal, and practical issues. For this reason, Meta set up its own Oversight Board.

\subsection{The Efficacy of the Disruption}
\noindent The disruption was more effective than previous DDoS attacks on the forum, as observed from our datasets. Yet the impact, although considerable, was short-lived. While part of the activity was shifted to Telegram, half of the core members returned quickly after the forum recovered. And while most casual users were shaken off, others turned up to replace them. Cutting forum activity and users by half might be a success if the goal of the campaign is just to hurt the forum, but if the objective was to ``drop the forum'', it has clearly failed. There is a lack of data on real-world harassment caused by forum members, such as online complaints or police reports, so we are unable to measure if the campaign had any effect in mitigating the physical and mental harm inflicted on people offline.

\kiwifarms suffered further DDoS attacks and interruption after our study period but it managed to recover quickly, at some point reaching the same activity level as before the interruption. It then moved primarily to the dark web in May-July 2023. The forum operator has shown commitment and persistence despite much disruption by DDoS attacks and infrastructure providers. He attempted to get \kiwifarms back online on the clearnet in late July 2023 under a new domain kiwifarms.pl, protected by their in-house DDoS mitigation system \kiwiflare, but the clearnet version appears to be unstable.

One lesson is that while repeatedly disrupting digital infrastructure might significantly lessen the activity of online communities, it may just displace them, which has been also noted in previous work~\cite{han2022infrastructure}. Campaigners can also get bored after a few weeks, while the disrupted community is more determined to recover their gathering place. As with the re-emergence and relocation of extremist forums like \eightchan and \dailystormer, \kiwifarms is now back online. This supports the argument that truly disrupting online active platforms can be very challenging, much like the short-term impact of shutting down cybercrime marketplaces~\cite{soska2015measuring}, DDoS-for-hire services~\cite{collier2019booting,kopp2019ddos}, and other security threats combining efforts of both law enforcement and industry interventions such as botnets and fraudulent ad networks when botmaster is capable of momentarily deploying new modules to counteract the takedown~\cite{nadji2013beheading, pearce2014characterizing}. Deplatforming alone may be insufficient to disperse or suppress an unpleasant online community in the long term, even when concerted action is taken by a series of tech firms over several months. It may weaken a community for a while by fragmenting their traffic and activity, and scare away casual observers, but it may also make core group members even more determined and recruit newcomers via the Streisand effect, whereby attempts at censorship can be self-defeating~\cite{hutchings2016taking,chua2019identifying}. 

\subsection{Censorship versus Free Speech} 
\noindent One key factor may be whether a community has capable and motivated defenders who can continue to fight back by restoring disrupted services, or whether they can be somehow disabled, whether through arrest, deterrence or exhaustion. This holds whether the defenders are forum operators or distributed volunteers. So under what circumstances might law enforcement take decisive action to decapitate an online forum, as the FBI did for example with the notorious \raidforums~\cite{raidforumtakedown} and \breachforums~\cite{breachforumtakedown}?

If some of a forum's members break the law, are they a dissident organisation with a few bad actors, or a terrorist group that should be hunted down? Many troublesome organisations do attract hot-headed young members, from animal-rights activists, climate-change protesters through to trade union organisers do occasionally fall foul of the law. But whether they are labelled as terrorists or extremists is often a political matter. People may prefer censoring harmful misinformation~\cite{kozyreva2023resolving}, yet taking down a website on which a whole community relies will often be hard to defend as a proportionate and necessary law-enforcement action. The threat of legal action can be countered by the operator denouncing whatever specific crimes were complained of. In this case, the \kiwifarms founder denounced SWAT attacks and other blatant criminality~\cite{kiwifarmsprinciple}. Indeed, a competent provocateur will stop just short of the point at which their actions will call down a vigorous police response. 

The free speech protected by the US First Amendment~\cite{schauer1991first} is in clear tension with the security of harassment victims. The Supreme Court has over time established tests to determine what speech is protected and what is not, including clear and present danger~\cite{clearandpresentdangertest}, a sole tendency to incite or cause illegal activity~\cite{badtendencytest}, preferred freedoms~\cite{preferredfreedomsdoctrine1,preferredfreedomsdoctrine2}, and compelling state interest~\cite{compellingstateinteresttest}; however, the line drawn between them is not always clear-cut. Other countries are more restrictive, with France and Germany banning Nazi symbolism and Turkey banning material disrespectful of Mustafa Kemal Atat\"urk. In the debates over the Online Safety Bill currently before the UK Parliament, the Government at one point proposed to ban `legal but harmful' speech online, while not making these speech acts unlawful face-to-face~\cite{anderson2023onlinesafetybill}. These proposals related to websites encouraging eating disorders or self-harm. Following the tragic suicide of a teenage girl~\cite{britishteensuicide}, tech firms are under pressure to censor such material in the UK using their terms of service or by tweaking their recommendation algorithms.

There are additional implications in taking down platforms whose content is harmful but not explicitly illegal. Requiring firms to do this, as was proposed in the Online Safety Bill, will drastically expand online content regulation. The UK legislation hands the censor's power to the head of Ofcom, the broadcast regulator, who is a political appointee. It will predictably lead to overblocking and invite abuse of power by government officials or big tech firms, who may suppress legitimate voices or dissenting opinions. There is an obvious risk of individuals or groups being unfairly targeted for political or ideological reasons.

\subsection{The Role of Industry in Content Moderation}
\noindent The rapid increase of cybercrime-as-a-service throughout the 2010s makes attacks easier than ever. A teenager with as little as \$10 can use a DDoS-for-hire service to knock your website offline~\cite{hutchings2016exploring}, so controversial websites depend on the grace and favour of a large hosting company or a specialist DDoS prevention contractor. This is just one aspect of a broader trend in tech: that the Internet is becoming more centralised around a small number of big firms, ranging from online social platforms, hosting companies, transit networks, to service providers and exchange points~\cite{mirrlees2021gafam}. While some do provide moderation tools that are favoured by content creators~\cite{thomas2022s}, some claim to be committed to fighting hate, harassment, and abuse yet are disproportionately responsible for serving online bad content~\cite{han2022infrastructure}, and the effort they put into the fight is variable~\cite{konikoff2021gatekeepers,heslep2021mapping}. Content moderation has recently shifted to the infrastructure layer~\cite{busch2022regulating}; now that activists have pressured infrastructure providers to act as content moderators, policymakers will be tempted too. Some may stand up to political or social pressure, because moderation is both expensive and difficult, but others may fold from time to time because of political pressure or legal compulsion. This would undermine the end-to-end principle of the Internet, as enshrined for example in COPA s 230 in the USA and in the EU's Net Neutrality Law~\cite{eunetneutrality}. 

Private companies must comply and remove illegal content from their infrastructure when directed to do so by a court order. However, deplatforming \kiwifarms or any other customers does not violate the principle of free speech. It is essentially a contractual matter; they have the right to cease their support for a website that violates their policies. Infrastructure providers may occasionally need to work expediently with law enforcement in the case of an imminent threat to life. Most providers have worked out ways of doing this, but the mechanisms can be too sluggish. \cloudflare attempted to collaborate with law enforcement to sort out the case of \kiwifarms, yet the process could not keep up with the escalating threats and it ended up taking unilateral action, relying on its terms of service~\cite{cloudflarebloackskiwifarms}. In an ideal world, we would have an international legal framework for taking down websites that host illegal content or that promote crime; unfortunately, this framework does not exist.

The Budapest Convention~\cite{seger2016budapest} criminalises some material on which all states agree, such as child sex abuse images, but even there the boundaries are contested~\cite{anderson2022chat}. Online drug markets such as \silkroad and \hansamarket have been taken down because of other laws -- drug laws -- that also enjoy international standardisation and collaboration. Copyright infringement also gets the attention of international treaties and coordinated action by tech majors, though civil law plays a greater role here than criminal law. Then there is material about which some states feel strongly but others do not; `one man's freedom fighter is another man's terrorist'. And then there's a vast swamp of fake news, animal cruelty, conspiracy theories, and other material that many find unpleasant or distressing, and which social networks moderate for the comfort of both their users and their advertisers. Legislators occasionally call for better policing of some of this content.

\subsection{Policy Implications}
\noindent Content moderation has become a political, policy, and public concern~\cite{gillespie2023expanding,alizadeh2022content}. The UK Online Safety Bill proposes a new regulator who will be able to apply for a court order mandating that tech firms disrupt an objectionable online activity~\cite{anderson2023onlinesafetybill}. One might imagine Ofcom deciding to take down \kiwifarms if their target had been a resident of Britain rather than Canada, and going to the various tech firms that were involved in the disruption we describe here, serving them one after another with orders signed by a judge in the High Court in London. Even if all the companies were to comply, rather than appealing or just ignoring the court altogether, it is hard to see how such an operation could be anything like as swift, coordinated or effective as the action taken on their own initiative by tech companies that we describe here. Where the censor's OODA loop -- the process by which it can observe, orient, decide and act -- involves a government agency assessing the effects of each intervention and then going to court to order the next one, the time constant would stretch from hours to months. And in any case, government interventions in this field are often significant but rather short-lived~\cite{collier2019booting,kopp2019ddos}. 

One factor contributing to the resilience of \kiwifarms is the technical competence of the forum owner. He has consistently and capably dealt with DDoS attacks on the forum, maintaining its codebase after xenForo stopped their licence, upgrading server hardware and network capability, and developing in-house DDoS protection mechanisms. Deplatforming can be more effective if the maintainer of a blatantly illegal website can be arrested and jailed (or otherwise incapacitated), as happened with \silkroad. With a forum like \kiwifarms, whose operator has denounced criminal acts perpetrated via his infrastructure~\cite{kiwifarmsprinciple}, the criminal-law option may simply not be available. The art of being a provocateur includes stopping just short of the point at which an aggressive criminal-law response would follow. This exposes the limits of civil-law remedies and voluntary action by platforms. 

Previous work has also explored why governments are less able to take down bad sites than private actors~\cite{hutchings2016taking}; that work analysed single websites with clearly illegal content, such as those hosting malware, phishing lures or sex-abuse images. This study shows why taking down an active community is likely to be even harder. Even when several tech firms roll their sleeves up and try to suppress a community some of whose members have indulged in crime and against whom there is an industry consensus, the net effect may be modest at best. Our case study may be the best result that could be expected for online censorship, but it only cut the users, posts, threads and traffic by about half. Our findings suggest that using content moderation law to suppress an unpleasant online community may be very challenging.

\subsection{Limitations and Future Work}
\noindent Measuring the link between physical harassment and \kiwifarms, as well as the cost of actual harm caused by the forum members to real-world victims, would be a valuable contribution. However, we lack ground-truth data about real-life events, which cannot be solely observed from forum discussions. Investigating doxxing-related posts that share real-victim information would be a good start, but the main challenge is validating data posted by untrusted users at scale, in the absence of a robust way to identify users. Our forum data is limited in studying user migration from \kiwifarms to its competitor \lolcowfarms as pseudonyms are unavailable on \lolcowfarms, so it is unclear if some \kiwifarms members have shifted there.

Our data scrapers are running in near real time, but there is still a chance of missing messages that are posted but removed swiftly thereafter. We expect the number of such missing messages to be relatively small. More insights can be revealed from private or protected posts as people can be more extreme when posting in private. However, we choose not to analyse them due to potential harm, legality, and ethical issues. \kiwifarms is now back online, and may well succeed in maintaining its accessibility on the clearnet. We will continue to monitor it, and extend our measurement to the more recent incidents in a follow-up report.

\section{Conclusion} \label{sec:conclusion}
\noindent Online communities may not only act as a discussion place but provide mutual support for members who share common values. For some, it may be where they hang out; for others, it may become part of their identity. Legislators who propose to ban an online community might consider precedents such as Britain's ban on Provisional Sinn F\'ein from 1988--94 due to its support for the Provisional IRA during the Troubles, or the bans on the Muslim Brotherhood enacted by various Arab regimes.\footnote{During the Sinn F\'ein ban, it was illegal to transmit the voice or image of their spokesmen in Britain, so the BBC and other TV stations employed actors to read the words of Gerry Adams and Martin McGuinness.} Declaring a community to be illegal and thus forcing it underground may foster paranoid worldviews, increase signals associated with toxicity and radicalisation~\cite{horta2021platform,ali2021understanding} and have many other unintended consequences. The \kiwifarms disruption, which involved a substantial concerted effort by the industry, is perhaps the best outcome that could be expected even if the censor were agile, competent and persistent. Yet this has demonstrated that merely attempting to deplatform an active standalone online community is not enough to deal effectively with hate and harassment, especially as the attempt failed to arrest, exhaust, or otherwise incapacitate the forum's maintainer. 

We believe the harm and threats associated with online hate communities may justify action despite the right to free speech. But within the framework of the EU and the Council of Europe which is based on the European Convention on Human Rights, such action will have to be justified as proportionate, necessary and in accordance with the law. It is unlikely that taking down a whole community or arresting its maintainer because of a crime committed by a single member can be proportionate. For a takedown to be justified as necessary, it must also be effective, and this case study shows how high a bar that could be. For a takedown to be in accordance with the law, it cannot simply be a response to public pressure. There must be a law or regulation that determines predictably whether a specific piece of content is illegal, and a judge or other neutral finder of fact would have to be involved.

The last time a Labour government won power in Britain, it won on a promise to be `Tough on Crime, and Tough on the Causes of Crime'. Some scholars of online abuse are now coming to a similar conclusion that the issue may demand a more nuanced approach~\cite{kumar2023understanding,vu2023extremebb}: as well as the targeted removal of content that passes an objective threshold of illegality, the private sector and governments should collaborate to combine takedowns with measures such as education and psycho-social support~\cite{onlineextremism}. And where the illegality involves violence, it is even more vital to work with local police forces and social workers rather than just attacking the online symptoms~\cite{anderson2022chat}. 

There are multiple research programmes and field experiments to effectively detox young men from misogynistic attitudes, whether in youth clubs and other small groups, at the scale of schools, or even by gamifying the identification of propaganda that promotes hate. But most countries still lack a unifying strategy for violence reduction~\cite{bates2021men}. In both the US and the UK, for example, while incel-related violence against women falls under the formal definition of terrorism, it is excluded from police counterterrorism practice, and the politicisation of misogyny has made this a tussle space in which political leaders and police chiefs have difficulty in taking effective action. In turbulent debates, policymakers should first ask which tools are likely to work, and it is in this context that we offer the present case study.

\section*{Acknowledgments}
\noindent We thank the anonymous reviewers and the shepherd for their insightful and constructive feedback. We are grateful to Richard Clayton, Alastair R. Beresford, Yi Ting Chua, Ben Collier, Tina Marjanov, Konstantinos Ioannidis, Daniel R. Thomas, and Ilia Shumailov for their invaluable comments. This work is supported by the European Research Council (ERC) under the European Union's Horizon 2020 research and innovation programme (grant agreement No 949127).
\bibliographystyle{IEEEtran}
\bibliography{main}

\newpage
\appendices
\section{Meta-Review}

\subsection{Summary}
This paper examines the impact of large-scale industry disruption on the online harassment forum \kiwifarms, as well as its competitor \lolcowfarms. The authors use a variety of measurement techniques to show a net reduction of activity on the forum.

\subsection{Scientific Contributions}
\begin{itemize}
\item Independent confirmation of important results with limited prior research.
\item Provides a new data set for public use.
\item Provides a valuable step forward in an established field.
\end{itemize}

\subsection{Reasons for Acceptance}
\begin{enumerate}
\item This paper provides a valuable step forward in the field of harassment measurement prevention by confirming important prior results. This paper examines deplatforming on an internet-wide scale, rather than focusing on one social network -- a limitation of most prior work.
\item This paper provides a new data set for public use. On request, the authors will provide a very detailed dataset of forum discussions with metadata, Telegram chats, web analytics, and relevant tweets, allowing independent confirmation and future research on harassment sites.
\end{enumerate}

\subsection{Noteworthy Concerns}
\begin{enumerate}
\item There is a significant lack of information about how the qualitative analysis of public announcements and press releases was conducted, which makes evaluation of that analysis challenging - details about reliability and how the coding process was done would be useful.
\item The discussion does a good job of describing why deplatforming is hard, but does not offer much in the way of suggestions for making this problem easier beyond arresting the people responsible.
\end{enumerate}
\end{document}